\documentclass[sigconf]{acmart}

\setcopyright{acmcopyright}
\copyrightyear{2022}
\acmYear{2022}
\acmDOI{XXX.XXX/XXX.XXX}

\acmConference[SIGMOD '22]{SIGMOD '22: ACM SIGMOD/PODS International Conference on Management of Data}{June 20–25, 2022}{Philadelphia, PA, USA}
\acmBooktitle{SIGMOD '22: ACM SIGMOD/PODS International Conference on Management of Data,
June 20–25, 2022, Philadelphia, PA, USA}
\acmPrice{15.00}
\acmISBN{XXX-XXX-XXX-XXX-XXX/18/06}
   
\usepackage{graphicx}
\usepackage[labelfont=bf]{caption}
\usepackage{makecell}
\usepackage{pifont}
\usepackage{soul}
\usepackage{multirow,array}
\usepackage{tablefootnote}
\usepackage{array, boldline, booktabs}
\usepackage[inline]{enumitem}
\usepackage{xspace}
\usepackage{colortbl}
\usepackage{hhline}
\usepackage{lipsum,graphicx,subcaption}
\hypersetup{breaklinks=true}
\let\oldding\ding% Store old \ding in \oldding
\renewcommand{\ding}[2][1]{\scalebox{#1}{\oldding{#2}}}% 

\usepackage{titlesec}
\titlespacing*{\section}{0pt}{2ex}{2ex}
\titlespacing*{\subsection}{0pt}{1ex}{1ex}
\usepackage{lipsum}

\newcommand\blfootnote[1]{%
  \begingroup
  \renewcommand\thefootnote{}\footnote{#1}%
  \addtocounter{footnote}{-1}%
  \endgroup
}

\usepackage{listings}
\usepackage{xcolor}

\definecolor{cmcolor}{rgb}{0.169, 0.2, 0.51}
\definecolor{codekey}{rgb}{0.216,0.49,0.13}
\definecolor{codecomment}{rgb}{0.31,0.49,0.50}
\definecolor{codegray}{rgb}{0.4,0.4,0.4}
\definecolor{codefunction}{rgb}{0,0.13,0.96}
\definecolor{codeconstant}{rgb}{0.92,0.2,0.137}

\newcommand{\showComments}{yes}
\newcommand{\note}[2]{
    \ifthenelse{\equal{\showComments}{yes}}{\textcolor{#1}{#2}}{}
}

\newcommand{\sysname} {\textsc{Sherman}\xspace}

\newcommand{\rsend} {\texttt{RDMA\_SEND}\xspace}
\newcommand{\rrecv} {\texttt{RDMA\_RECV}\xspace}
\newcommand{\rwrite} {\texttt{RDMA\_WRITE}\xspace}
\newcommand{\rread} {\texttt{RDMA\_READ}\xspace}
\newcommand{\ratomic} {\texttt{RDMA\_ATOMIC}\xspace}
\newcommand{\rcas} {\texttt{RDMA\_CAS}\xspace}
\newcommand{\rfaa} {\texttt{RDMA\_FAA} \xspace}

\newcommand{\btree} {$\rm B^+Tree$\xspace}

\newcommand{\linksym}{\color{codecomment}\mmapsto}

\definecolor{linecolor}{rgb}{0.38,0.157,0.404}
\newcommand{\codeline}[1]{\hyperref[fig:code]{\color{linecolor}{#1}}}
\newcommand{\codelock}[1]{\hyperref[code:lock]{\color{linecolor}{#1}}}
\newcommand{\codewrite}[1]{\hyperref[code:write]{\color{linecolor}{#1}}}
\newcommand{\codesearch}[1]{\hyperref[code:search]{\color{linecolor}{#1}}}

\newcommand{\eg}{e.g.,\xspace}
\newcommand{\ie}{i.e.,\xspace}

\begin{document}

\title{Sherman: A Write-Optimized Distributed B$^+$Tree Index on Disaggregated Memory}

\author{Qing Wang}
\affiliation{%
  \institution{Tsinghua University}
  \city{Beijing}
  \country{China}
}

\author{Youyou Lu}
\affiliation{%
  \institution{Tsinghua University}
  \city{Beijing}
  \country{China}
}

\author{Jiwu Shu}
\affiliation{%
  \institution{Tsinghua University}
  \city{Beijing}
  \country{China}
}

\begin{abstract}
% \sysname is a write-optimized distributed \btree index on RDMA networks
% that performs all index operations via one-sided verbs.
% \sysname leverages unexplored RDMA hardware features 
% and incorporates efficient software techniques,
% to reduce round trips, accelerate locks,
% and mitigate write amplification.

Memory disaggregation architecture physically separates CPU and memory into independent components, which are connected via high-speed RDMA networks, greatly improving resource utilization of databases.
However, such an architecture poses unique challenges to data indexing in databases due to limited RDMA semantics and near-zero computation power at memory-side.
Existing indexes supporting disaggregated memory either suffer from low write performance, or require hardware modification.

This paper presents \sysname, a write-optimized distributed \btree index on disaggregated memory that delivers high performance with commodity RDMA NICs.
\sysname combines RDMA hardware features and RDMA-friendly software techniques 
to boost index write performance from three angles.
First, to reduce round trips,
\sysname coalesces dependent RDMA commands by leveraging in-order delivery property of RDMA.
Second, to accelerate concurrent accesses, \sysname introduces a hierarchical lock that exploits on-chip memory of RDMA NICs.
Finally, to mitigate write amplification, 
\sysname tailors the data structure layout of \btree with a two-level version mechanism.
Our evaluation shows that,
\sysname is one order of magnitude faster 
in terms of both throughput and 99th percentile latency
on typical write-intensive workloads,
compared with state-of-the-art designs.

\end{abstract} 

% \begin{CCSXML}
%     <ccs2012>
%      <concept>
%       <concept_id>10010520.10010553.10010562</concept_id>
%       <concept_desc>Computer systems organization~Embedded systems</concept_desc>
%       <concept_significance>500</concept_significance>
%      </concept>
%      <concept>
%       <concept_id>10010520.10010575.10010755</concept_id>
%       <concept_desc>Computer systems organization~Redundancy</concept_desc>
%       <concept_significance>300</concept_significance>
%      </concept>
%      <concept>
%       <concept_id>10010520.10010553.10010554</concept_id>
%       <concept_desc>Computer systems organization~Robotics</concept_desc>
%       <concept_significance>100</concept_significance>
%      </concept>
%      <concept>
%       <concept_id>10003033.10003083.10003095</concept_id>
%       <concept_desc>Networks~Network reliability</concept_desc>
%       <concept_significance>100</concept_significance>
%      </concept>
%     </ccs2012>
% \end{CCSXML}
    
%     \ccsdesc[500]{Computer systems organization~Embedded systems}
%     \ccsdesc[300]{Computer systems organization~Redundancy}
%     \ccsdesc{Computer systems organization~Robotics}
%     \ccsdesc[100]{Networks~Network reliability}

% \keywords{indexing, memory disaggregation, RDMA}

\settopmatter{printfolios=true}
\settopmatter{printacmref=false}
\maketitle

\blfootnote{This is the pre-print version of our SIGMOD'22 paper.}

\section{Introduction}
\label{sec:intro}

% Dis trend
The popularity of in-memory databases (e.g., SAP HANA~\cite{HANA}) and in-memory computing (e.g., Spark~\cite{Spark})
catalyzes ever-increasing demands for memory in modern datacenters.
However, datacenters today suffer from low memory utilization (< 65\%)~\cite{EuroSys20Borg, IWQoS19Trace,OSDI19LegoOS, maruf2021memtrade}, which results from imbalanced memory usages across a sea of servers.
In response, academia and industry are working towards a new hardware architecture called 
\emph{memory disaggregation}, where CPU and memory are \emph{physically} separated into two network-attached components -- compute servers and memory servers~\cite{ISCA09Memory, OSDI19LegoOS, ASPLOS21DisMem,OSDI16Network, MICRO20ThymesisFlow,keeton2015machine, VLDB20Disaggreation, OSDI20Semeru, SOSP21Mind, APSys21DiLOS, asplos22Clio, SuperNIC, farview,FAST22FORD}.
With memory disaggregation, CPU and memory can scale independently and
different applications share a global disaggregated memory pool efficiently.

Since almost all CPUs are assembled on compute servers under the memory disaggregation architecture, 
memory servers have near-zero computation power,
which highlights the challenge that how compute servers access disaggregated memory residing on memory servers.
Fortunately, RDMA (Remote Direct Memory Access),
a fast network technique, allows compute servers to \emph{directly access} disaggregated memory 
unmediated by memory servers'  computation power  with a single-digit-microsecond level latency,
becoming an essential building block of memory disaggregation architecture~\cite{OSDI19LegoOS,VLDB20Disaggreation, OSDI16Network, ASPLOS21DisMem}.

In this paper, we explore how to design a high-performance tree index, a key pillar of database systems, on disaggregated memory.
We first revisit existing RDMA-based tree indexes and examine their applicability on disaggregated memory.
Several RDMA-based tree indexes rely on remote procedure calls (RPCs) to handle write operations
\footnote{In this paper, we refer \emph{lookup} and \emph{range query}
as read operations,
and refer \emph{insertion} (include update existing keys)
and \emph{deletion} as write operations.}~\cite{Cell, FaRMOpacity};
they are ill-suited for disaggregated memory due to near-zero computation power of memory servers.
For tree indexes that can be deployed on disaggregated memory, 
they also have some critical limitations:
\ding[1.2]{192}
Some indexes using RDMA one-sided verbs for all index operation~\cite{FG}
(we call it \emph{one-sided approach});
they can deliver high performance for read operations, but suffer from low throughput and high latency in terms of write operations, 
especially in high-contention scenarios (< 0.4 Mops with 
\textasciitilde20ms tail latency, \S\ref{sub:tradeoff}).
\ding[1.2]{193}
Other indexes bake write operations into SmartNICs or 
customized hardware~\cite{HTTree}, 
which brings high TCO (total cost of ownership)  
and cannot be deployed in datacenters in a large scale immediately.

Our goal is designing a tree index on disaggregated memory
that can 
\emph{deliver high performance for both read and write operations with commodity RDMA NICs}.
To this end, we further analyze
what makes one-sided approach inefficient in write operations,
and find out three major causes.
First, due to limited semantics of one-sided verbs,
modifying an index node 
(\eg tree node in \btree)
always requires multiple 
round trips (\ie lock, read, write, unlock),
inducing high latency and further making 
conflicting requests more likely to be blocked.
Second,
the locks used for resolving write-write conflicts are slow
and experience performance collapse under 
high-contention scenarios.
This is because 
\ding[1.2]{192} 
at the hardware level, 
NICs adopt expensive concurrency control to ensure atomicity between RDMA atomic commands, 
where each command needs two PCIe transactions.
\ding[1.2]{193}
at the software level, 
such locks always trigger unnecessary retries, which 
consumes RDMA IOPS, and do not 
provide fairness, which leads to high tail latency.
Third,
the layout of index data structure incurs severe write amplification.
Due to coarse-grained consistency check mechanisms
(\eg using checksum to protect a whole tree node),
a small piece of modification will result in large-sized write-back
across the network.

% Why

% We 

% Index is  many systems like 
% databases, file systems and key-value stores.
% Recently,
% RDMA-based indexes, 
% which are enabled by 
% high performance and direct access 
% (\ie using one-sided verbs 
% % --- write, read and atomic instructions --- 
% to bypass remote CPUs) features of RDMA network,
% gain considerable interest in both academia and industry
% ~\cite{FaRM, FaRMOpacity, DrTM, Cell, FG}.
% Different from traditional ones,
% RDMA-based indexes can host an unprecedented amount of data 
% by scaling beyond a single machine boundary in a CPU efficient way.

% Although various optimizations have successfully 
% improved performance of read operations

% for RDMA-based indexes~\cite{FaRM, DrTM, Cell},
% \eg caching and lock-free search,
% write operations are overlooked.
% Specifically,
% existing RDMA-based indexes 

Motivated by the above analysis,
we propose \sysname
\footnote{The largest tree in the world is a giant sequoia called 
General Sherman, which stands in California's Sequoia National Park.},
a write-optimized distributed \btree index on disaggregated memory.
The key idea of \sysname is \emph{combining RDMA hardware features 
and RDMA-friendly software techniques
to reduce round trips, accelerate lock operations, and mitigate write amplification}.
\sysname spreads \btree nodes across a set of memory servers,
and compute servers perform all index operations via RDMA one-sided verbs purely.
\sysname uses a classic approach for concurrency control:
 lock-free search with versions to resolve read-write conflicts and exclusive
locks to resolve write-write conflicts~\cite{FG, Eurosys12MassTree}.

To reduce round trips, 
\sysname introduces a command combination technique.
Based on the observation that RDMA NICs already provide in-order delivery property,
this technique allows client threads to issue dependent RDMA commands 
(\eg write-back and lock release) simultaneously,
letting NICs at memory servers reflect them into disaggregated memory in order.

To accelerate lock operations, we design a \emph{hierarchical on-chip lock (HOCL)}
for \sysname.
HOCL is structured into two parts: global lock tables on memory servers,
and local lock tables on compute servers.
Global lock tables and local lock tables coordinate conflicting lock requests
between compute servers and within a compute server, respectively.
Global lock tables are stored in the on-chip memory of RDMA NICs,
thus eliminating PCIe transactions of memory servers
and further delivering extremely high
throughput for RDMA atomic commands (\textasciitilde110 Mops).
Within a compute server,
before trying to acquire a global lock on memory servers,
a thread must acquire the associated local lock,
so as to avoid a large amount of unnecessary remote retries.
Moreover, by adopting wait queues,
local lock tables improve fairness between conflicting lock requests.
Based on local lock tables, 
a thread can hand over its acquired lock
to another thread directly, 
reducing at least one round trip for acquiring global locks.

To mitigate write amplification,
\sysname tailors the leaf node layout of \btree.
First, entries in leaf node are \emph{unsorted},
so as to eschew shift operations upon insertion/deletion.
Second, 
to support lock-free search while avoiding write amplification,
we introduce a \emph{two-level version mechanism}.
In addition to using a pair of \emph{node-level versions} to detect
the inconsistency of the whole leaf node,
we embed a pair of \emph{entry-level versions} into each entry,
which ensures entry-level integrity.
For insertion/deletion operations without split/merging events, 
only entry-sized data is written back,
thus saving network bandwidth and 
making the most of the extremely high IOPS of small RDMA messages.

We evaluate \sysname using a set of benchmarks.
Under write-intensive workloads,
\sysname achieves much better performance than FG~\cite{FG}, a state-of-the-art distributed \btree supporting disaggregated memory.
Specifically, in common skewed workloads, 
\sysname delivers one order of magnitude performance improvement in terms of both throughput and 99th percentile latency.
Meanwhile, a \sysname tree can be accessed concurrently by more than 500 client threads,
while providing peak throughput stably.
For read-intensive workloads (i.e., 95\% read operations), 
\sysname exhibits slightly higher throughput with 25\% lower 99th percentile latency.

% read 

% \vspace{0.2cm}
\noindent
\textbf{Contributions.}
The main contributions of this paper are:

\begin{itemize}[labelsep=5pt]
    \item An analysis of existing tree indexes on disaggregated memory,
    demonstrating that 
    the inefficiency of write operations in the one-sided approach
    stems from excessive round trips,
    slow synchronization primitives and write amplification (\S\ref{sec:moti}).
    
    % \vspace{0.1cm}
    \item The design and implementation of \sysname,
    a write-optimized \btree index on disaggregated memory,
    which boosts write performance by combining RDMA hardware features and 
    RDMA-friendly software techniques (\S\ref{sec:design}).

    % \vspace{0.1cm}
    \item A set of evaluations that demonstrate the high performance of \sysname
    under different workloads (\S\ref{sec:eval}).

  \end{itemize}

\section{Background}
\label{sec:bg}

\begin{figure}
	
	\centering
	\includegraphics[width=0.9\linewidth]{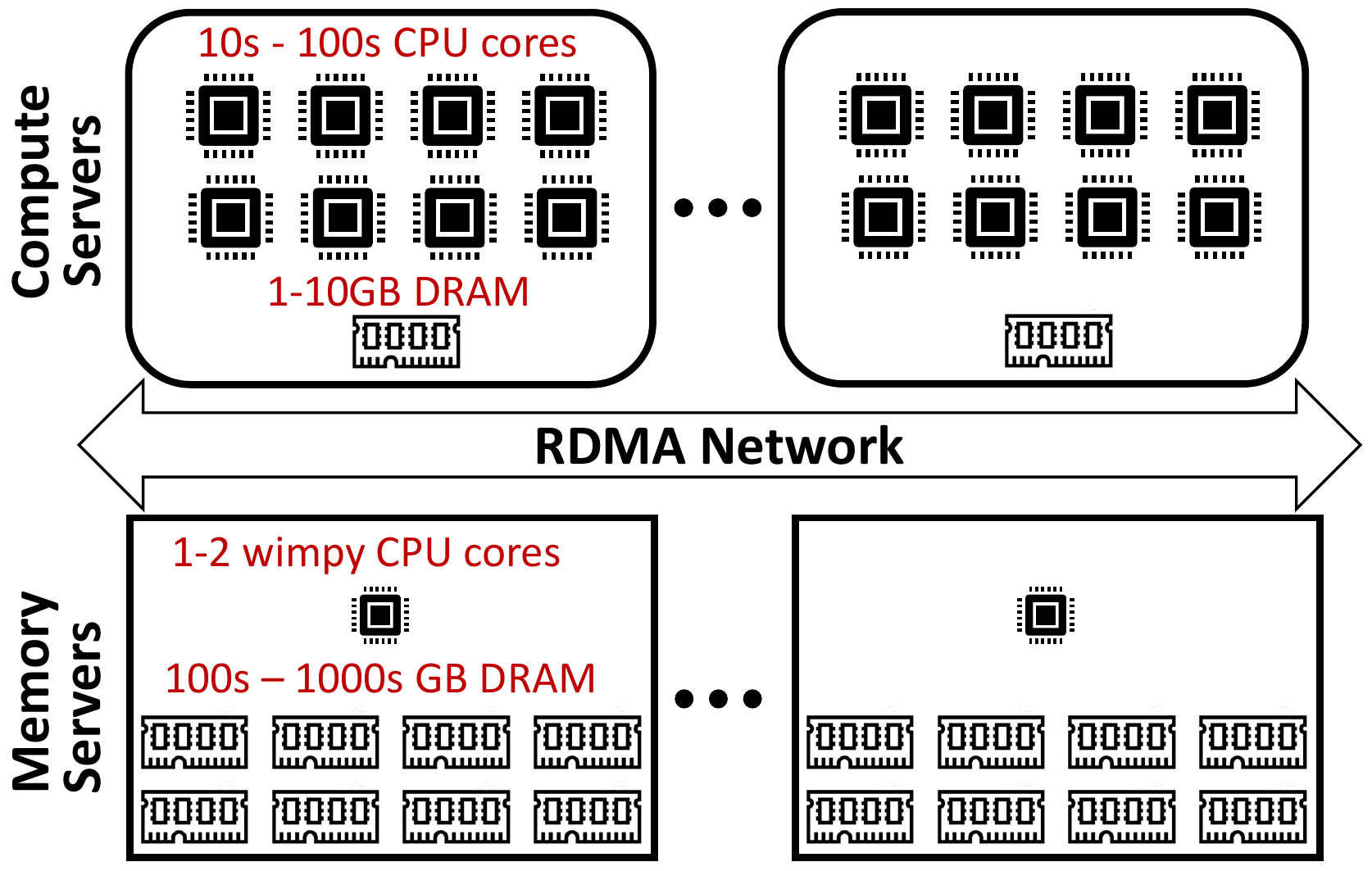}
	
	% \vspace{-0.1in}
	\caption{Architecture of Memory Disaggregation. }
	% \vspace{-0.1in}
	
	\label{fig:dis-mem}
\end{figure}

In this section, we provide the background on memory disaggregation (\S\ref{subsec:memdis}) and RDMA network (\S\ref{subsec:rdma}) briefly.

\subsection{Memory Disaggregation}
\label{subsec:memdis}

Traditional datacenters pack CPU and memory into the same hardware units (\ie monolithic servers),
leading to low memory utilization (< 65\%) ~\cite{OSDI19LegoOS, IWQoS19Trace, EuroSys20Borg}
and further increasing the TCO (total cost of ownership) of datacenters.
To attack this problem, academia and industry are exploring a new hardware architecture called \emph{memory disaggregation}~\cite{ISCA09Memory, OSDI16Network, OSDI19LegoOS, ANCS16SQL, VLDB20Disaggreation}, which is shown in Figure~\ref{fig:dis-mem}.
In such an architecture, CPU and memory are physically separated into two different hardware units: 
compute servers (CSs) and memory servers (MSs).
CSs own a mass of CPU cores (10s - 100s), but MSs host high-volume memory (100s - 1000s GB) with near-zero computation power.
CPUs in CSs can directly access the disaggregated memory in MSs via high-speed RDMA networks (\S\ref{subsec:rdma}). 
With memory disaggregation, CPU and memory can scale independently 
and applications can pack resources in a more flexible manner, boosting resource utilization significantly. 

To reduce remote accesses from CSs to MSs, 
CSs are always equipped with a small piece of memory as the local cache (1 - 10 GB).
Moreover, MSs owns a small set of wimpy CPU cores (1 - 2) to support lightweight management tasks,
such as network connection management and disaggregated memory allocation.

\subsection{RDMA Network}
\label{subsec:rdma}

RDMA network is the key enabler of memory disaggregation architecture.
RDMA is increasingly popular in modern datacenters due to high bandwidth (\eg 100Gbps) and
low latency ($\le 2\mu$s)~\cite{FaRM,Sigcomm16RdmaAtScale}.
RDMA provides two types of verbs, 
namely \emph{two-sided verbs} and \emph{one-sided verbs},
to applications.
Two-sided verbs --- \rsend and \rrecv --- 
are the same as traditional Linux socket interface:
the sender generates messages via \rsend,
and the receiver pre-posts \rrecv commands 
as well as processes incoming messages.
One-sided verbs, \ie \rwrite, \rread and
\ratomic (\rfaa and \rcas), operate directly on remote memory
without involving the CPUs of receivers.
The direct-access feature of one-sided verbs makes memory servers with near-zero computation power possible.

RDMA hosts communicate via queue pairs (QPs).
A QP consists of a send queue and a receive queue,
and a completion queue (CQ) is associated with the QP.
A sender performs an RDMA command by 
posting the request on the send queue.
On completion, the sender's NIC writes a completion entry into 
the CQ,
and the sender can know it by polling the CQ.
RDMA supports three transport types: reliable connected (RC), unreliable connected (UC) and unreliable datagram (UD).
\sysname uses RC, since it supports all one-sided verbs and is reliable;
yet, RC requires one-to-one connections between QPs 
(\ie a QP can only communicate with one QP).

Sparked by high performance and new one-sided verbs of RDMA,
there is an active line of research in RDMA-based indexes\cite{FaRM,FaRMOpacity,DrTM, Cell,HTTree, FG}.
In this paper, we focus on distributed \btree index on disaggregated memory 
due to its elasticity
and support for range query, but our ideas can be applied to other kinds of indexes.
%\vspace{-0.15in}
\section{Motivation}
\label{sec:moti}

Designing a high-performance distributed tree index on disaggregated memory poses unique challenges.
In this section, we first revisit two existing approaches
--- using one-sided verbs purely and extending RDMA interfaces --- and reveal their respective issues (\S\ref{sub:tradeoff}).
Then, we analyze why using one-sided verbs is slow, which motivates the design of \sysname (\S\ref{subsec:moti_slow}).

\subsection{Existing Approaches}
\label{sub:tradeoff}

With near-zero computation power at MS-side, 
we cannot delegate index operations to CPUs of MSs via remote procedure calls (RPCs),
which is the main difference between memory disaggregation and traditional architectures.
Existing work enables efficient lookup operations via lock-free search and caching mechanisms~\cite{FaRM, Cell, FG, OSDI20XStore}, leading to a single \rread in the ideal situation (\ie cache hit).
However, write operations are in a quandary due to their complex semantics;
specifically, there are two avenues to design write operations,
each of which has its own issues.

\begin{table}[!t]
	\begin{center}		
	
		\resizebox{0.48\textwidth}{!}{
			\begin{tabular}{|c|c|r|r|r|r|}
				\hhline{~~----}
				\multicolumn{2}{c|}{}  &  \multicolumn{2}{c|}{\textbf{read-intensive}} & \multicolumn{2}{c|}{\textbf{write-intensive}} \\
				\hhline{~~----}
			\multicolumn{2}{c|}{}  &  \textbf{uniform} & 
			\textbf{skew}
			 & \textbf{uniform} & \textbf{skew} \\
			\hline
			
			\multicolumn{2}{|c|}{\textbf{Throughput (Mop/s)}}  &  31.8  & 32.9
			&  18.7 &  \cellcolor[gray]{0.95} 0.34  \\
		   \hline
		   \multirow{3}{*}{\textbf{Latency ($\mu$s)}} & \textbf{50th} & 4.9 & 4.7 & 9.5 &  \cellcolor[gray]{0.95} 10\\
		   \hhline{~-----}
		   & \textbf{90th}  & 6.4 & 6.2 & 14.3 & \cellcolor[gray]{0.95} 68.7\\
		   \hhline{~-----}
		   & \textbf{99th} & 14.9 & 15.3 & 19 & \cellcolor[gray]{0.95} 19890\\
		   \hline
			\end{tabular}
				}
		% \vspace{0.5cm}
		\caption{Index performance in one-sided approach (100 Gbps ConnectX-5 NICs, 8 MSs, 8 CSs with 176 client threads, 8/8-byte key/value, 1 billion key space).
		% The query:update ratio is 95:5 in the \emph{read-intensive} workload,
		% and 50:50 in the \emph{write-intensive} one.
		% Key popularity in \emph{skew} setting follows a Zipfan distribution of skewness 
		The performance collapses under \emph{write-intensive and skew} setting.
		}

		\label{table:moti}	

		% \vspace{-0.5cm}
		
	\end{center}
\end{table}

\begin{table*}[!t]
	\begin{center}		
	
		% \resizebox{0.7\textwidth}{!}{
			\begin{tabular}{|c|c|c|c|c|c|}
              
                \cline{2-6}
				\multicolumn{1}{c|}{} & \textbf{Cell}~\cite{Cell} & \textbf{FaRM-Tree}~\cite{FaRMOpacity} & \textbf{FG}~\cite{FG} & \textbf{HT-Tree}~\cite{HTTree} & \textbf{\sysname} \\
				\hline
                \textbf{Read Performance}& Medium & High & Medium & High & High\\
                \hline
                \textbf{Write Performance }& Medium & High & Low & High & High\\
				\hline				
				\textbf{No Hardware Modification}& \ding{51} & \ding{51} & \ding{51} & \ding{55} & \ding{51}\\
                \hline
				\textbf{Support Disaggregated Memory}& \ding{55} & \ding{55} & \ding{51} & \ding{51} & \ding{51}\\
				\hline

			\end{tabular}
				% }
				\vspace{0.2cm}
		\caption{A comparison of RDMA-based distributed tree indexes.
		All indexes use one-sided verbs for read operations.
		HT-Tree is a hybrid index combining the hash table and tree,
		and the remaining four are B$^+$Tree structure.
		Cell executes write operations via RPCs; it only 
		caches tree nodes near the root (at least four node levels above the leaf node level), so all read/write operations
		need four round trips at least.
		FaRM-Tree implements write operations by exploiting transactions,
		which in turn uses RPCs; it caches all internal tree nodes. 
		FG leverages one-sided verbs to realize write operations,
		and does not use any caching mechanism.
		HT-Tree is a conceptual design without implementation,
		which relies on hardware extensions.
		}
		\label{table:comparison}	
		% \vspace{0.1in}
		
	\end{center}
\end{table*}

\subsubsection{Using One-sided Verbs Purely.}
FG~\cite{FG}, a distributed RDMA-based \btree, is the only one tree index 
that completely leverages one-sided verbs to perform index write operations
(so it can be deployed on disaggregated memory).
Specifically, FG uses a B-link tree structure~\cite{BLinkTree} and distributes tree nodes across different servers.
For write operations, it adopts a lock-based approach, where tree node modification is protected by RDMA-based spinning locks;
the locks leverage \rcas for lock acquisition and \rfaa for lock release.
For read operation, FG follows a lock-free search scheme,
where threads fetch tree nodes via \rread without holding locks and then check the consistency using checksum\footnote{Private communication with authors of FG~\cite{FG}.}.
FG does not adopt any caching mechanism.

We conduct an experiment to evaluate FG's performance.
% We demonstrate the inefficiency of the existing design
% by conducting an experiment.
Since FG is not open-source, 
we implement it from scratch;
we also cache internal tree nodes to reduce remote accesses.
Table~\ref{table:moti} shows the result,
and we make two observations.
First, 
in read-intensive workload (95\% lookup and 5\% insert/update),
FG can achieve high throughput (> 30Mops)
% \footnote{Note that the results is not peak throughput due to lacking client threads.
% By attaching 4 coroutines in a client thread, we can get about 2$\times$ higher throughput.})
and low 99th percentile latency (< 16$\mu$s).
Second,
in write-intensive workload (50\% lookup and 50\% insert/update),
FG delivers 18 Mops with 19$\mu$s tail latency under the \emph{uniform} setting;
yet, its performance collapses in case of skew setting,
where the key popularity follows a Zipfian popularity with skewness 0.99:
the throughput is only 0.34 Mops and corresponding tail latency
is close to 20ms.
This result is unacceptable,
considering the two characteristics of today's datacenter workloads:
\begin{enumerate*}[label=(\itshape\roman*\upshape)]
	\item It is well-known that
	data access patterns in the real-world follow a power-law distribution (\ie Zipfian distribution)~\cite{YCSB,FacebookWorkloads,HotNet14Imbalance},
	where a small number of hot items receive frequent access.
	\item More and more applications exhibit write-intensive workloads,
	such as graph computation~\cite{OSDI12PowerGraph}, 
	parameter servers~\cite{OSDI14PS}, and data warehousing system~\cite{NSDI20Snowflake}.
\end{enumerate*}
The inefficiency of write operations comes from 
excessive round trips, slow synchronization primitives
and write amplification, which we will  
elaborate in ~\S\ref{subsec:moti_slow}.

% \noindent
% \textbf{Use RPCs.}
% Some RDMA-based indexes abandon one-sided verbs,
% the key features of RDMA,
% for write operations;
% instead, they delegate write operations to
% remote CPUs via RPCs~\cite{FaRMOpacity,Cell,FaRM}.
% This approach raises three issues.
% First, 
% we have to reserve a considerable amount of 
% remote CPU cores to serve write operations;
% for example, with 256-byte packet size,
% saturating a 100Gbps RDMA NIC needs at least 6 CPU cores.
% Such reservation wastes the CPU cycles 
% that can be re-assigned to other tasks like transaction processing.
% Second, 
% in high-speed RDMA network,
% the common practice to receive short-term RPC requests is spin-polling.
% However, spin-polling suffers from imbalance, head-of-line-blocking
% and scalability issues~\cite{ASPLOS19RPCValet,NSDI19Shinjuku,SOCC19Spin};
% further, it wastes energy~\cite{EnergyComputing}
% and speeds up CPU cores aging~\cite{DSN12ExtraTime}.
% Third,
% in promising future datacenter architecture, namely 
% hardware resource disaggregation~\cite{OSDI16Disaggregation, OSDI19LegoOS, ATC20pDPM}, 
% servers equipping massive memory 
% have limited processing power,
% so RPCs are impractical.

\subsubsection{Extending RDMA Interfaces.}
Another approach to designing tree indexes on disaggregated memory 
is extending RDMA interfaces.
This approach offloads index write operations into memory servers' NICs via
SmartNICs or other customized hardware~\cite{NSDI14AzureSmartNIC, SIGCOMM19IPipe, HTTree, SOSP21PRISM, SOSP21Xenic, SOSP21LineFS, EuroSys20STROM}.
For example, 
by exploiting interface extensions 
(\ie indirect addressing and notification),
researchers propose HT-Tree (without implementation)~\cite{HTTree},
a hybrid index combining the hash table and tree.
Yet, compared with commodity RDMA NICs,
SmartNICs come at the price of 
\emph{higher TCO (total cost of ownership)}
and \emph{lower performance}.
More specifically,
SmartNICs have much higher market price (\textasciitilde5$\times$) than that of
commodity RDMA NICs at present~\cite{ATC20pDPM}.
As for performance, 
at 100Gbps network environment,
StRoM~\cite{EuroSys20STROM},
the state-of-the-art RDMA extensions using FPGA,
has 2$\times$ higher \rread/\rwrite round-trip latency (4$\mu$s)
against commodity RDMA NICs ($\le 2\mu$s).

Table~\ref{table:comparison} shows a comparison of RDMA-based distributed tree indexes.
Among them, Cell~\cite{Cell} and FaRM-Tree~\cite{FaRMOpacity} use RPCs for write operations, so they cannot be deployed on disaggregated memory.
FG~\cite{FG} suffers from low write performance and HT-Tree~\cite{HTTree} needs hardware modification.
\sysname aims to achieve high performance (for both read and write operations) with commodity RDMA NICs on disaggregated memory.

% The \textbf{goal} of \sysname is to break the three-way tradeoff;
% designers of RDMA-based distributed tree indexes
% confront a three-way tradeoff
% between \ul{\emph{high performance}},
%  \ul{\emph{remote CPU bypass}},
% and \ul{\emph{no hardware modification}}.

% in other words, 

\subsection{Why One-sided Approach is Slow?}
\label{subsec:moti_slow}

Using analysis and experiments, 
we show the inefficiency of write operations using 
one-sided verbs stems from \emph{excessive round trips}, 
\emph{slow synchronization primitives} and \emph{write amplification}.

\subsubsection{Excessive Round Trips.}
The most obvious cause of slow write operations
is excessive round trips.
For example,
when modifying a tree node (not consider node split/merging),
a client thread needs \texttt{4} round-trips:
\ding[1.2]{192}acquiring associated exclusive lock,
\ding[1.2]{193}reading the tree node,
\ding[1.2]{194}writing back the modified tree node
\ding[1.2]{195}and finally releasing the lock.
Excessive round trips negatively
impact write performance in two aspects.
First, 
the latency of a single write operation 
is proportional to the number of round trips.
Second,
more round trips lead to the longer critical path,
so conflicting write operations (\ie requests targeting at the same tree node)
are more likely to be blocked,
degrading the concurrency performance.

\begin{figure}[t]
	
	\centering
	\includegraphics[width=0.85\linewidth]{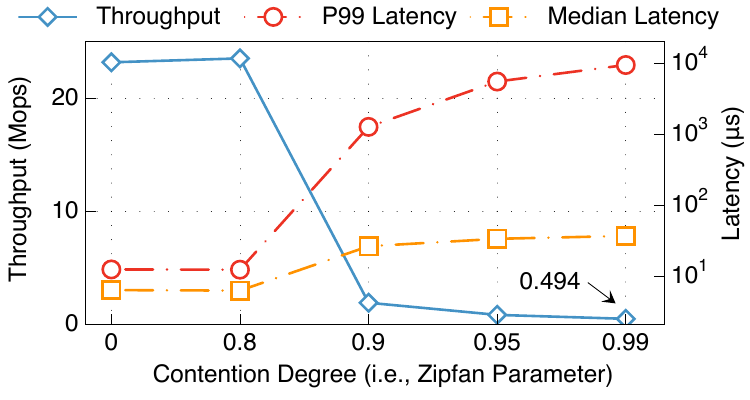}
	
	% \vspace{-0.14in}
	\caption{Performance of RDMA-based exclusive locks
	(ConnectX-5 NICs, 100 Gbps). 
	The system experiences performance collapse
	under high-contention 
    settings.}
	% \vspace{0.2in}
	
	\label{fig:moti_lock}

\end{figure}

\subsubsection{Slow Synchronization Primitives.}

RDMA-based locks used in these indexes cannot provide 
sustainable performance under a variety of workloads.
We conduct an experiment to present performance issues of such locks.
In the experiment, 
154 threads across 7 CSc acquire/release
10240 locks residing in an MS;
we choose the RDMA verbs used by FG~\cite{FG},
where \rcas for lock acquisition and 
\rfaa for release.
The access pattern follows Zipfian distribution
and we vary Zipfian parameter
to control the contention degree
(0 is uniform setting, and 0.99 is the most common 
real-world scenario).
Figure~\ref{fig:moti_lock} shows the result.
The system experiences performance collapse
in terms of throughput and latency under high-contention settings.
We attribute the performance collapse to the following three reasons.

\noindent
\textbf{Expensive in-NIC concurrency control.}
To guarantee correct atomicity semantic 
between \ratomic commands targeting the same addresses,
RDMA NICs adopt
an internal locking scheme~\cite{ATC16Guidelines}.
More specifically, 
a NIC maintains a certain number of buckets (\eg 4096),
and puts \ratomic commands having the same certain bits
in their destination addresses (\eg 12 LSBs) into the same bucket.
Commands in the same bucket are considered conflicting.
An \ratomic can not be executed until the previous conflicting commands
are finished.
Unfortunately, 
a single \ratomic needs two PCIe transactions:
\ding[1.2]{192} reading data from CPU memory into the NIC
and \ding[1.2]{193}writing back after modification
(\ding[1.2]{193} can be omitted in case of failed \rcas).
These PCIe transactions stretch the queue
time of conflicting \ratomic commands and thus 
degrade concurrent performance of \ratomic commands, 
especially in high-contention workloads.

\noindent
\textbf{Unnecessary retries.}
When failing to acquire a lock,
the client thread retries.
Different from locks designed for shared memory with coherent cache,
such retries in the RDMA environment require remote network accesses,
squandering limited throughput of NICs in both senders and receivers.
Retries can be eliminated via notification mechanisms,
where a client thread is notified when the lock it wants to acquire has been released.
Yet, notification needs remote CPU involvement, which is impractical on disaggregated memory,
or specialized hardware like programmable switches~\cite{SIGCOMM20NetLock}.

\noindent
\textbf{Lacking Fairness.}
Existing locks in RDMA-based indexes 
do not take into account fairness between conflicting lock acquisition,
thus starving some client requests and 
further inducing high tail latency.

\begin{figure}[t!]
	
	\centering
	\includegraphics[width=0.85\linewidth]{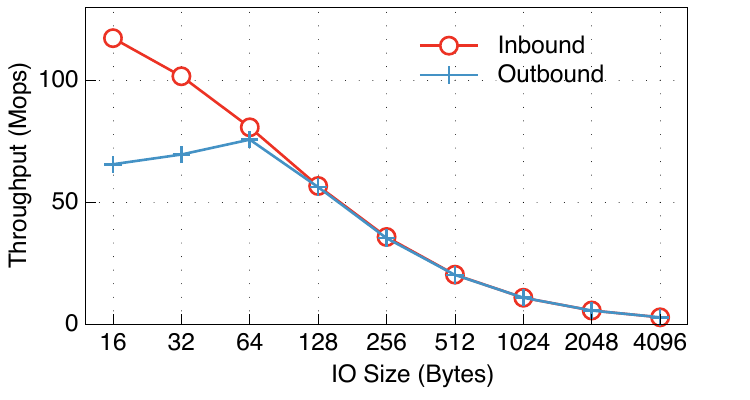}
	
	\vspace{-0.1in}
	\caption{Performance of \rwrite (ConnectX-5 NICs, 100 Gbps).
	\rwrite prefers small IO size.
	}
	% \vspace{-0.1in}
	
	\label{fig:moti_write}
\end{figure}

\subsubsection{Write Amplification.}
RDMA prefers small IO size, 
but indexes using one-sided verbs fail to exploit this feature.
Figure~\ref{fig:moti_write}
shows the throughput of outbound/inbound \rwrite under varied IO sizes\footnote{Outbound/inbound throughput:  
the throughput of sending/receiving \rwrite commands from a NIC.}.
When IO size is less than or equal to 128 bytes, 
the throughput is more than 50 Mops;
however, when IO size reaches to 256 bytes,
NICs's hardware bandwidth restricts throughput;
our result is consistent with previous studies~\cite{DrTM_H}.
Yet, existing RDMA-based tree indexes always trigger large-sized \rwrite,
suffering from write amplification.
This issue stems from two causes. 

\noindent
\textbf{Sorted layout.}
A \btree keeps entries in each tree node sorted,
to support traversal in internal nodes
and binary search in leaf nodes.
To ensure this property,
when an entry is inserted/deleted to/from a node,
all the entries on the right side of 
the insertion/deletion position need to be shifted.
The shift operations cause extra data to be written via \rwrite.

\begin{figure}[t!]
	
	\centering
	\includegraphics[width=0.9\linewidth]{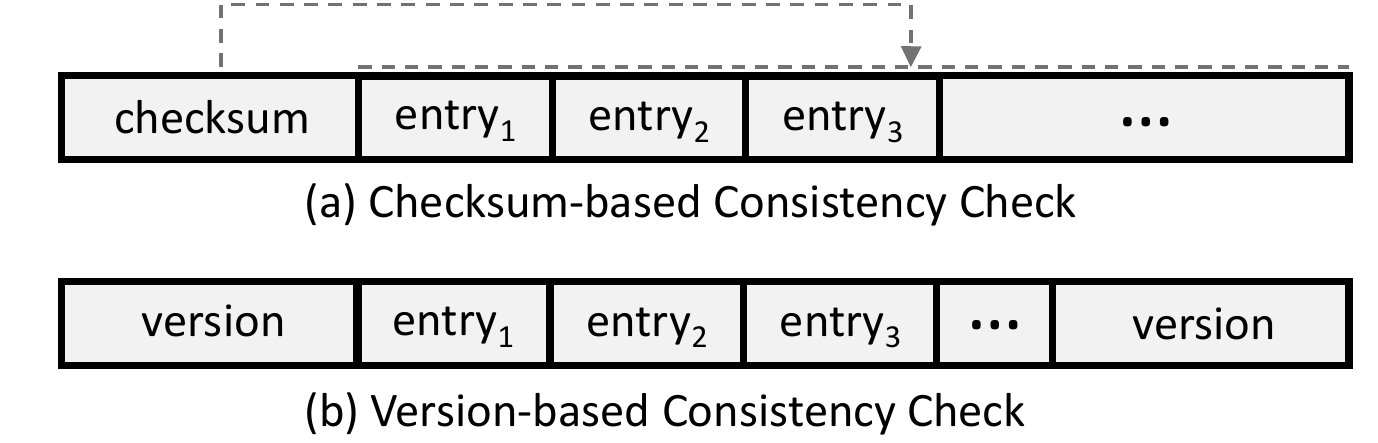}
	
	\vspace{-0.14in}
	\caption{Two mechanisms to detect inconsistency of tree nodes.
	Both of them lead to write amplification.
	}
	\vspace{-0.1in}
	
	\label{fig:moti_read}
\end{figure}

\noindent
\textbf{Coarse-grained consistency check.}
In order to read tree nodes in a lock-free way 
and detect incomplete data due to ongoing writes,
%  through issuing only one \rread,
two main consistency check mechanisms are proposed.
In the first mechanism (Figure~\ref{fig:moti_read}(a)),
each node includes a checksum covering the whole node area (except the checksum itself)~\cite{ATC13Pilaf,FG};
the checksum is re-calculated when modifying the corresponding node,
and is verified when reading the node.
The other mechanism, namely version-based consistency check (Figure~\ref{fig:moti_read}(b)),
stores a version number at the start and end of each node~\cite{Cell};
when modifying a node via \rwrite, the corresponding two version are incremented; a node's content obtained via \rread is consistent only when
the two versions are the same\footnote{Like Cell~\cite{Cell} and NAM-DB~\cite{VLDB17NAMDB}, 
we observe that the NIC reads data in increasing address order,
so we omit per-cacheline version mechanisms used by FaRM~\cite{FaRM}.
}.
Since the granularity of the above two mechanisms is tree node,
any modification to part of the node area requires to write back the 
whole node 
(include the metadata, \eg checksum and version),
leading to severe write amplification.

% \pagebreak

\section{Design}
\label{sec:design}

\vspace{0.2cm}
Motivated by our observations about root causes of 
inefficient write operations (\S\ref{subsec:moti_slow}),
we design \sysname, 
a write-optimized distributed \btree index on disaggregated memory.
In this section, 
we begin by presenting our design principles (\S\ref{sub:principle}),
proceed to give an overview of \sysname (\S\ref{sub:overview}), 
and then describe our key techniques
(\S\ref{sub:hlock}-\S\ref{sub:cc}).
Finally, we conduct a discussion (\S\ref{sub:discuss}).
% , which centering on
% the core algorithms depicted in Figure~\ref{fig:code}.

\subsection{Design Principles}
\label{sub:principle}

\sysname boosts the performance of write operations 
by reducing round trips, 
accelerating lock operations,
and mitigating write amplification.
It follows three core design principles.
 
\noindent
\textbf{1) Seeking solutions from hardware first.}
There are hiding features of RDMA NICs that pose
opportunities for index design.
In \sysname, 
we utilize
\ding[1.2]{192}
on-chip memory exposed by NICs to accelerate lock operations,
and 
\ding[1.2] {193} in-order delivery guarantee of RC queue pairs
to reduce round trips.

\noindent
\textbf{2) Applying CS-side optimization when possible.}
In the current multicore era, 
each compute server (CS) always launches a mass of client threads (10s - 100s)
to manipulate an index concurrently,
leaving a lot of design space for CS-side optimization.
Within a CS,
we leverage local lock tables to coordinate conflicting lock requests between threads,
and locks can be handed over from a thread to another quickly.
Besides, \sysname maintains CS-side index cache to reduce network accesses for tree traversal.

\noindent
\textbf{3) Tailoring data structure layout to improve RDMA friendliness.}
Disaggregated memory has unique profile:
it is accessed via network (i.e., RDMA) rather than cache-coherent memory bus.
Thus, it is necessary to tailor data structure layout of indexes to reap RDMA's full potential.
To this end, 
\sysname \ding[1.2]{192} separates locks from the tree structure, to put 
them into NICs' on-chip memory, and
\ding[1.2]{193} adopts entry-level versions as well as unsorted leaf nodes,
to reduce IO size of \rwrite commands.

\subsection{Overview}
\label{sub:overview}

\begin{figure}[t!]
	
  \centering
  \includegraphics[width=\linewidth]{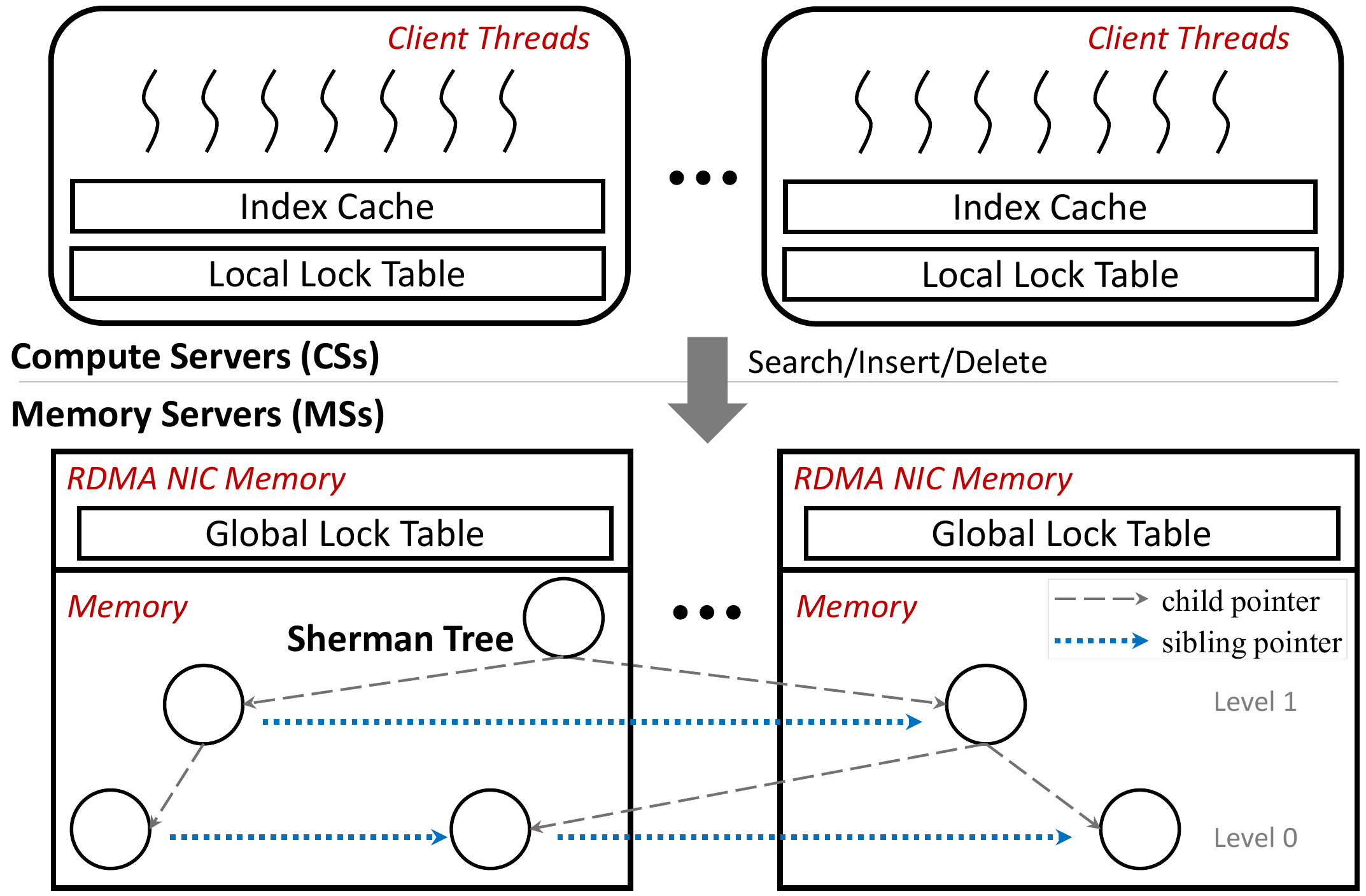}
  
  % \vspace{-0.14in}
  \caption{\sysname's architecture and interactions.}
  \vspace{-0.1in}
  
  \label{fig:overview}
\end{figure}

Figure~\ref{fig:overview} shows 
the overall architecture and interactions of \sysname,
which consists of a set of memory servers (MSs) and compute servers (CSs).
MSs are equipped with massive memory where our \sysname tree resides.
CSs run client threads that manipulate \sysname 
via specific interfaces (\ie lookup, range query, insert and delete).
Each MS and CS is equipped with RDMA NICs for network communication.
Since the number of CPU cores continues to increase
~\cite{DAC07ThousandCore, VLDB14ThousandCore,DaMoN20ThousandCore},
we assume that there are always a host of client threads running within a CS
(dozens or even hundreds).
These client threads cooperatively 
execute system services (e.g., transaction processing),
which needs \sysname for data indexing.

% Note that CNs and SNs can co-locate 
% at the same physical machines~\cite{FaRM,DrTM}.

\subsubsection{B$^+$Tree Structure.}
\sysname is a \btree, where values are stored in leaf nodes.
We record a sibling pointer for every leaf node and internal node as in the B-link tree~\cite{BLinkTree}.
Client threads can always reach a targeted node by following these sibling pointers in the presence of node split/merging, thus supporting concurrent operations efficiently.
Every pointer in \sysname (\ie child/sibling pointers)
is 64-bit, which includes two parts:
16-bit MS unique identifier and 48-bit memory address within corresponding MS.

\subsubsection{Concurrency Control.}
\sysname adopts lock-based write operations and lock-free read operations.

\noindent
\textbf{Write-write conflicts}. 
\sysname uses \emph{node-grained} exclusive locks to resolve write-write conflicts:
before modifying a tree node, the client thread must acquire the associated exclusive lock.
These exclusive locks are hierarchical: 
\emph{local lock tables} at CS-side
and \emph{global lock tables} in on-chip memory of MSs' NICs
(\S\ref{sub:hlock}). Such a hierarchical structure avoids PCIe transactions at MS-side, reduces remote retries, and improves fairness.

\noindent
\textbf{Read-write conflicts}. 
\sysname supports lock-free search, which leverages \rread
to fetch data residing in MSs without holding any lock.
Moreover,
\sysname uses versions to detect inconsistent data caused by concurrent writes.
However, different from traditional mechanisms that use node-level versions,
\sysname proposes a two-level version mechanism, 
which combines entry-level and node-level versions, 
to mitigate write amplification (\S\ref{sub:tlversion}).

\subsubsection{Cache Mechanism.}
To reduce remote accesses in the tree traversal,
\sysname adopts a cache mechanism.
Each CS maintains an \emph{index cache},
which only makes two types of internal nodes' copies:
\ding{182} nodes above the leaf nodes (level 1 in Figure~\ref{fig:overview}),
and \ding{183} the highest two level of nodes (including root).
A client thread firstly searches type \ding{182} cache.
On hit, it fetches the targeted leaf node directly from MSs;
otherwise, it searches type \ding{183} cache and
then traverses to leaf nodes via remote accesses.
The type \ding{183} is always cached;
the type \ding{182} is structured as a lock-free skiplist,
and applies power-of-two-choices~\cite{PowerOfTwo} for evicting:
selecting two cached nodes randomly 
and evicting the one least recently used.
The index cache never induces data consistency issues for two reasons.
First, internal nodes in \btree only contain location information,
and the real data stored in leaf nodes are always accessed by RDMA commands.
Second, in every node, we store a lower bound and upper bound for the set of keys that can appear in it (\ie \emph{fence keys}~\cite{BTreeLocking}), 
and we also store the level of the node (leaf nodes are level 0). 
When fetching a node from MSs,
we check whether fence keys and node level are legal;
if not, we invalidate the cache entry steering us to the wrong node 
and retry.

\subsubsection{Memory Management.}
In each MS, we reserve a dedicated \emph{memory thread} to manage disaggregated memory.
The memory thread divides memory residing in corresponding MS into fixed-length chunks (\ie 8MB).
Client threads use a two-stage memory allocation scheme to obtain memory from MSs.
A client thread first chooses an MS in a round-robin manner, and 
requests a free chunk from the MS's memory thread via RPCs.
Then, it allocates memory space for tree nodes locally within the chunk.
Such a two-stage scheme comes with two advantages.
First, it avoids network communication for most allocation operations.
Second, it mitigates processing overhead of memory thread significantly,
which is necessary considering the near-zero computation power of MSs (i.e., 1-2 wimpy CPU cores).
Note that the round-robin allocation may cause imbalanced accesses across MSs; we
leave addressing this issue for future work.

For deallocation, 
since all memory space allocated in \sysname is fixed-length (\ie node size)
and each tree node owns its internal verification information 
(\ie version, fence keys and node level),
we do not need complex garbage collection strategies 
(\eg epoch-based reclamation~\cite{FaRM}).
Instead, client threads only need to clear 
a \emph{free} bit in a tree node before deallocating it;
later requests that fetch the garbage node will 
realize the node has been freed.

\begin{figure}[!t]
  \centering

  \begin{lstlisting}[firstnumber=1,,
      mathescape=true, columns=fullflexible, xleftmargin=0.2cm]
  MAX_LOCK_PRE_MS = 131072 # the number of locks in each MS
  MAX_DEPTH = 4 # maximum number of consecutive handovers
  
  def HOCL_Lock(addr):
  idx = calculate_$\texttt{hash}$(addr) % MAX_LOCK_PRE_MS
  l = LLT[addr.ms_id][idx]
  if lock(l) == False: # get locks in LLT first
    handover = False
    # wait queue ensures first-come-first-served fairness
    l.wait_queue.push(&handover)
    while &handover != l.wait_queue.head():
      continue
    lock(l) # get locks in LLT first
    l.wait_queue.pop()
    if handover == True: # be handed over a lock
      return
  # get locks in GLT finally: RDMA_CAS(addr, compare, swap)
  while $\rcas$(&GTL[addr.ms_id][idx], 0, cs_id) == False:
    continue
  
  def HOCL_Unlock(addr, combine_list):
  idx = calculate_$\texttt{hash}$(addr) % MAX_LOCK_PRE_MS
  l = LLT[addr.ms_id][idx]
  handover = l.wait_queue.head()
  if handover != None && (++l.handover_depth) <= MAX_DEPTH:
    *handover = true # hand over lock to another thread
  else:
    l.handover_depth = 0
    # collect an RDMA_WRITE command for lock (in GLT) release
    # command's details: write 0 into 16-bit GLT[addr.ms_id][idx]
    combine_list.push{(0, &GLT[addr.ms_id][idx], 16bit)}
  
  unlock(l) # release locks in LLT
  $\rwrite$ combine_list # posting combined commands in list
\end{lstlisting}

\caption{Pseudo-code of hierarchical on-chip lock (HOCL).
\texttt{combine\_list} is used to collect commands of \rwrite that can be issued together (\S\ref{sub:cc}).}

\label{code:lock}

\end{figure}

\subsection{Hierarchical On-Chip Lock}

\label{sub:hlock}

\sysname proposes hierarchical on-chip lock (HOCL) to improve
concurrency performance.
HOCL leverages on-chip memory of NICs
to avoid PCIe transactions at MS-side;
it also maintains local locks at CS-side,
to form a hierarchical structure,  reduce retries and improve fairness.
Moreover, locks can be handed over between client threads
within the same CSs, thus saving at least one round trip.
Figure~\ref{code:lock} shows the pseudo-code of HOCL.

\noindent
\textbf{On-chip lock table.}
Current RDMA verbs support \emph{device memory programming}
~\cite{RDMADeviceMemory}.
Specifically, 
an RDMA NIC can expose a piece of its on-chip memory to the upper applications, which can be allocated and read/written by RDMA commands.
The on-chip memory eliminates PCIe transaction at receiver-side,
thus providing extremely high throughput (\textasciitilde110 Mops \rcas).
\sysname separates locks from tree nodes, 
and stores locks into on-chip memory at MS-side;
each tree node is in the \emph{same} MS as
the lock protecting it.
These locks in each MS are structured as an array,
namely \emph{global lock table} (GLT).
When locking a tree node,
the client thread first hashes the address of the tree node 
into an index number in the corresponding GLT
 (line \codelock{5} in Figure~\ref{code:lock}),
and then issues an \rcas command to the lock, which tries to change 
it from \texttt{0} to the 16-bit CS identifier atomically (line \codelock{18}).
For lock release, the client thread clears the lock via an \rwrite command
~\cite{DrTM_H} (line \codelock{31}, \codelock{34}).

An important consideration of GLT is 
the on-chip memory size.
In the NIC we use (i.e., ConnectX-5), 256KB on-chip memory is available.
To accommodate more locks, 
we make the granularity of \rcas finer (16 bits rather than 64 bits),
by applying an infrequently used RDMA verb called
\emph{masked compare and swap}~\cite{EnhancedAtomic},
which allows us to select a portion of 64-bit for \rcas operations.
Thus, an MS can maintain 131,072 locks in its GLT,
enabling extremely high concurrency, 
particularly considering that
we only lock at most one tree node at a time for a single write operation
~\cite{PODS85BLinkLocking}.
To the best of our knowledge,
\sysname is the first RDMA-based system that leverages on-chip memory of commodity RDMA NICs.
A concurrent work leverages NIC's on-chip memory to accelerate network functions~\cite{asplos22NICMem}.

\noindent
\textbf{Hierarchical structure.}
\sysname maintains a \emph{local lock table} (LLT) in each CS,
to coordinate conflicting lock requests within the same CSs.
The LLT stores a \emph{local lock} for each lock of all GLTs.
When a thread needs to lock a tree node, it first acquires the associated local lock in LLT
(lines \codelock{6}, \codelock{7}, \codelock{13}), and then acquires the associated lock in GLT;
thus, 
conflicting lock requests from the same CSs are queued on the LLT at CS-side,   
avoiding unnecessary remote retries and thus saving RDMA IOPS.
Moreover,
each local lock in LLT is associated with a \emph{wait queue}.
A thread that cannot acquire a local lock in LLT
pushes itself into the corresponding queue;
the thread can learn if its turn has arrived by checking 
whether it is at the head of the queue (lines \codelock{8}-\codelock{14}).
The queue provides first-come-first-served fairness
among threads within the same CSs.
For lock release, the thread first releases the lock in GLT and then the local lock in LLT.
Of note, an LLT consumes small local memory space of each CS;
in our implementation, each local lock is 8-byte, so an LLT uses \emph{n} MB space, where \emph{n} is the number of MSs (i.e., $n * 8 * 131072$ bytes).

\noindent
\textbf{Handover mechanism.}
The hierarchical structure of HOCL enables 
a handover mechanism:
handing over a lock from one client thread to another.
When releasing a lock, 
if a thread finds out the lock's wait queue is not empty,
it will hand over the lock to the one at the head of the wait queue
(lines \codelock{24}-\codelock{26}).
To avoid starving threads at other CSs, 
we limit the maximum number of consecutive handovers to \texttt{4}.
The thread that is handed over a lock no longer needs remote accesses
for acquiring the lock, thus saving at least one round trip
(lines \codelock{15}-\codelock{16}).

\begin{figure}[!t]
  \centering

  \begin{lstlisting}[firstnumber=1,,
      mathescape=true, columns=fullflexible, xleftmargin=0.2cm]
  NODE_SIZE # size of a tree node 
  ENTRY_SIZE # size of an entry
    
  def Sherman_insert(key, value):
    leaf_addr = index_cache.find(key) # search index cache

    HOCL_Lock(leaf_addr) # lock the targeted leaf node 
    node = $\rread$(leaf_addr, NODE_SIZE) # RDMA_READ(remote_addr, size);
    combine_list = {} # RDMA_WRITE task list: (buffer, addr, size)
  
    if $\exists$i,node.entry[i].k == key || node.entry[i].k == None: #update/insert
      # entry-level node modification
      node.entry[i].{k, v} = {key, value}
      node.entry[i].front_entry_ver++
      node.entry[i].rear_entry_ver++
      # collect a combined command for write-back of the entry in leaf node
      combine_list.push({node, leaf_addr+offset(entry[i]), ENTRY_SIZE})
    else # split
      sibling = malloc(NODE_SIZE) # local buffer
      sibling_addr = remote_alloc(NODE_SIZE) # allocate memory at MSs
      node.sort().move_half_entries_to(sibling) # sort and move
      sibling.{level, free} = {0, 0}
      link(node, sibling) # link sibling and list: ... node $\linksym$ sibling  ...
      update_fence_keys(node, sibling)
      # node-level node modification
      node.add(key, value)
      node.front_node_ver++
      node.rear_node_ver++
      if leaf_addr.ms_id == sibling_addr.ms_id:
        # collect a combined command for write-back of the sibling node
        combine_list.push({sibling, sibling_addr, NODE_SIZE})
      else:
        $\rwrite$(sibling, sibling_addr, NODE_SIZE)
      # collect a combined command for write-back of the leaf node
      combine_list.push({node, leaf_addr, NODE_SIZE}) 
    
    HOCL_Unlock(leaf_addr, combine_list) # unlock the targeted leaf
    if sibling != None: # update internal nodes
      insert_internal(sibling.entry[0].k, sibling_addr)\end{lstlisting}

  \vspace{-0.1cm}
  \caption{Pseudo-code of \sysname's insertion procedure.
  To simplify exposition, we omit the details on 1) traversing the tree in case of index cache miss and 2) following sibling pointers in case of concurrent node split/merging.}

  \label{code:write}
\end{figure}

\subsection{Two-Level Version}
\label{sub:tlversion}
To address the write amplification issue,
\sysname incorporates a \emph{two-level version mechanism}.
First, \sysname uses unsorted leaf nodes 
so that shift operations upon insertion/deletion can be avoided.
Unsorted leaf nodes complicate write operations in two aspects, 
as shown in Figure~\ref{code:write}:
\begin{enumerate*}[label=(\itshape\roman*\upshape)]
    \item When looking up a key,
    the client thread needs to traverse the entire targeted leaf node (line \codewrite{11}).
	\item Before splitting a leaf node,
	the client thread must sort the entries in it (line \codewrite{21}).
\end{enumerate*}
Given the microsecond level network latency,
the added overhead is slight.
Although previous work on persistent 
memory data structures leverages unsorted nodes
to reduce persistence overhead~\cite{Chen2011RethinkingDA,FAST15NVTree},
we are the first one that uses it to alleviate write amplification of indexes on disaggregated memory.

\begin{figure}[t!]
	
  \centering
  \includegraphics[width=\linewidth]{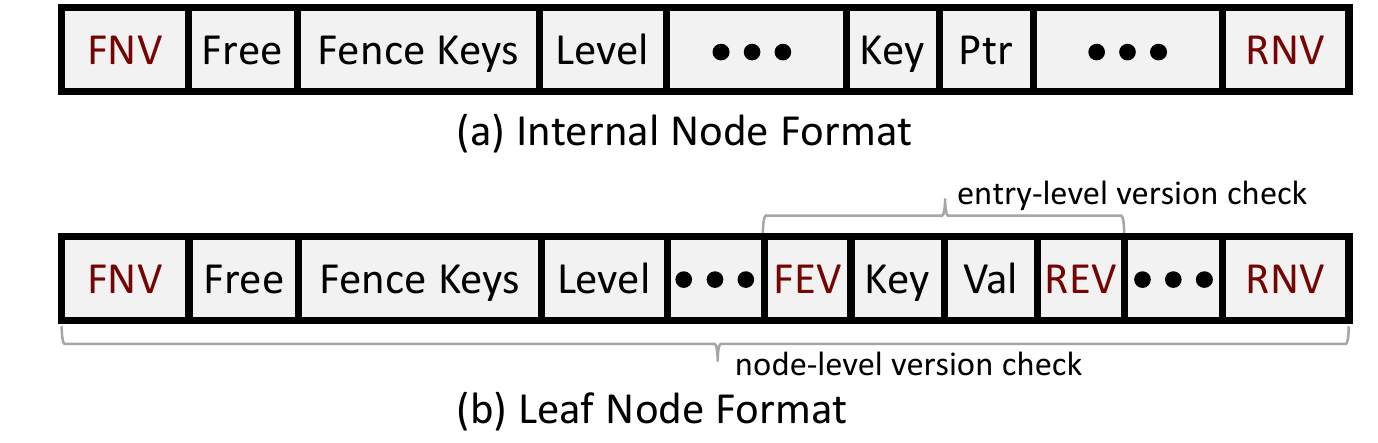}
  
  % \vspace{-0.14in}
  \caption{The format of internal nodes and leaf nodes in \sysname.
  \textbf{FNV/RNV} is 4-bit \underline{f}ront/\underline{r}ear
  \underline{n}ode \underline{v}ersion;
  \textbf{FEV/REV} is 4-bit \underline{f}ront/\underline{r}ear 
  \underline{e}ntry \underline{v}ersion.
  The entries in leaf nodes are unsorted.
  Entry-level versions are updated in case of insertion/deletion without
  split/merging, thus alleviating write amplification.
  Node-level versions are updated in case of node split/merging.
  }

  \label{fig:node}
\end{figure}

Second, \sysname introduces \emph{entry-level versions} to 
enable fine-grained consistency check, as shown in Figure~\ref{fig:node}.
Specifically, in leaf nodes, each entry is surrounded by 
a pair of 4-bit \emph{entry-level} versions (\ie FEV and REV).
In case of insertion without splitting,
the associated entry-levels versions are incremented 
and only the modified entry (includes FEV and REV) is written back via \rwrite
(lines \codewrite{13}-\codewrite{17} in Figure~\ref{code:write}),
thus evading write amplification.
Also, a pair of 4-bit \emph{node-level versions} (\ie FNV and RNV) 
is stored at the begin and end of each leaf node,
protecting the consistency at the node granularity.
When splitting a leaf node, the client thread increments associated FNV and RNV,
and writes back the whole node via \rwrite (lines \codewrite{26}-\codewrite{33}).
Since internal nodes have a much lower modification frequency 
than leaf nodes, 
their format is standard:
two node-level versions with a sorted layout.
The extra memory space occupied by entry-level versions is acceptable:
considering a \btree storing 8-byte key and 8-byte value, 
each key-value pair at leaf nodes needs extra 1-byte memory space for entry-level versions,
consuming about 6\% memory of all leaf nodes.

\noindent
\textbf{Lookup operation.}
Our two-level version mechanism makes the lookup procedure different from 
the standard one.
Figure~\ref{code:search} shows the pseudo-code of \sysname's lookup procedure.
After reading a leaf node from MSs,
a client thread compares the two node-level versions first;
mismatched versions indicate the read must be retried (lines \codesearch{6}-\codesearch{7}).
Then, the client thread locates the targeted entry
and compares the two associated entry-level versions;
if the comparison fails, the client thread needs to
re-read the leaf node via \rread (lines \codesearch{10}-\codesearch{11}).
Wraparounds of these 4-bit versions may
cause inconsistency to be undetected:
two versions match but one wraps around.
To handle wraparounds, 
we measure the time of each \rread command:
if it takes more than 8$\mu$s (\ie $2^4 \times 0.5$, where $0.5\mu s\ll$ the time of a single write operation),
a retry is needed.

\noindent
\textbf{Range query.}
For a range query, 
the client thread issues multiple 
\rread in parallel to fetch targeted leaf nodes,
and then checks leaf nodes' consistency in the same way as the lookup procedure.
Of note, like FG, \sysname does not guarantee that a range query is atomic with concurrent write operations.
If upper applications, e.g., transaction engine, need snapshot semantics for range queries,
they must use other mechanisms to avoid phantom problem~\cite{Slio}.

\noindent
\textbf{Delete operation.}
Delete operations experience a similar process as insert operations.
In case of no leaf node merging, a delete operation only needs to clear the targeted entry (i.e., set key to \emph{null}),
increase the associated entry-level versions, and write back them via \rwrite.    
When triggering node merging during a delete operation, \sysname uses node-level version to detect 
inconsistency.

\subsection{Command Combination}
\label{sub:cc}

To enforce ordering between dependent RDMA commands (\eg writing back a node and then releasing the lock),
existing RDMA-based indexes use an expensive approach:
issuing the following RDMA command only after receiving the acknowledgement of the preceding one~\cite{FG}.
Yet, we observe that RDMA already provides a strong ordering property
at the hardware level:
in a reliable connected (RC) queue pair,
\rwrite commands are transmitted in the order 
that they are posted,
and the NIC at receiver-side executes these commands in order
~\cite{IBSpec,VLDB19ActiveMemory}.
By leveraging this ordering property,
\sysname combines multiple \rwrite commands in a write operation,
so as to reduce round trips.

There are two cases that \sysname combines multiple \rwrite commands.
First, since a tree node and its associated lock 
co-locate at the same MS,
the write-back of tree node and lock release can be combined through a QP
(lines \codewrite{17}, \codewrite{35} in Figure~\ref{code:write} and
line \codelock{31} in Figure~\ref{code:lock}),
as opposed to issuing an unlock request after receiving 
acknowledgement of write-back; 
thus, one round trip is saved and the critical path shortens.
Second, when a node (we call it \texttt{A} here) splits,
we check whether the newly allocated sibling node belongs the same MS as \texttt{A};
if so, 
three \rwrite commands can be combined together:
\ding[1.2]{192} write-back of the sibling node,
\ding[1.2]{193} write-back of \texttt{A}
and \ding[1.2]{194} release of \texttt{A}'s lock
(line \codewrite{31} in Figure~\ref{code:write}).

Client threads issue these combined commands to the targeted MS
by posting a linked list of them in one call (line \codelock{34} in Figure~\ref{code:lock});
such a combination not only saves round trips,
but also reduces CPU usages and PCIe transactions.
In addition, we only mark the last one in the list as signaled
(\ie generate a completion entry to the corresponding CQ),
so NIC-initiated DMAs for writing completion entries
can be reduced~\cite{ATC16Guidelines}.

\begin{figure}[!t]
  \centering

  \begin{lstlisting}[firstnumber=1,,
      mathescape=true, columns=fullflexible, xleftmargin=0.2cm]
  def Sherman_lookup(key):
    leaf_addr = index_cache.find(key)
    retry:
      node = $\rread$(leaf_addr, sizeof(node))
      # node-level check
      if node.front_node_ver != node.rear_node_ver: 
        goto retry
      if $\exists$e, e in node && e.k == key: # find targeted entry
        # entry-level check
        if e.front_entry_ver != e.rear_entry_ver:
          goto retry
        return e.v
    return None  \end{lstlisting}
    \caption{Pseudo-code of \sysname's lookup procedure.}

    \label{code:search}
  \end{figure}

\begin{figure*}[!t]
  \centering

  \centering
  \includegraphics[width=\linewidth]{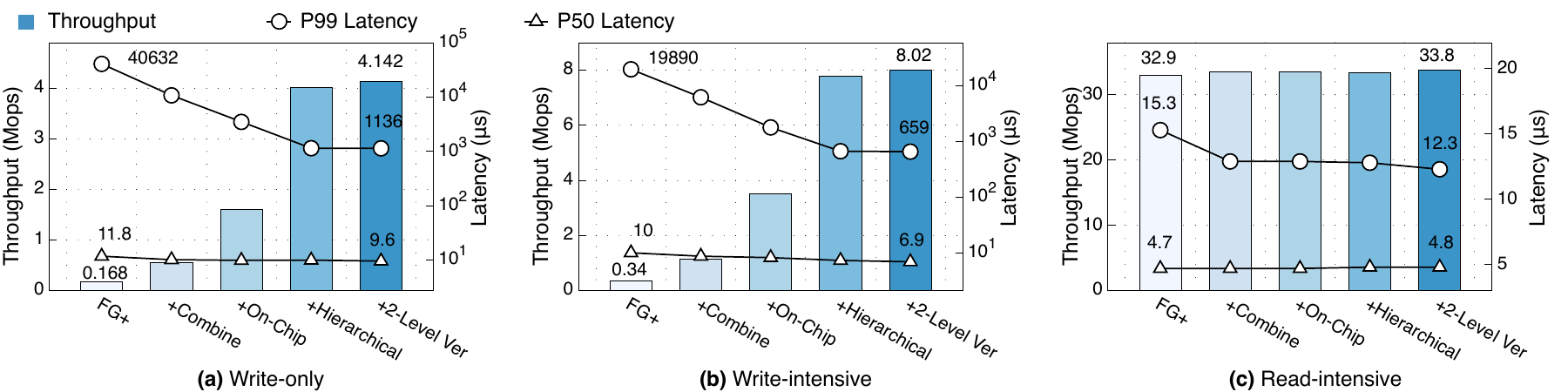}

  \caption{Contributions of techniques to performance
    (skewed workloads, skewness=0.99).}

  \label{fig:overall:skew}
\end{figure*}

  \subsection{Discussion}
  \label{sub:discuss}

\noindent
\textbf{Generality of RDMA features used by \sysname.}
\sysname leverages two RDMA hardware features: in-order property and on-chip memory.
The in-order property is defined in RDMA specification~\cite{IBSpec}.
The on-chip memory is supported in Mellanox's ConnectX-5 NICs (announced in 2016) and above~\cite{RDMADeviceMemory}.
Note that ConnectX-5 NICs and above are already widely used in datacenters; for example, Alibaba uses ConnectX-5 NICs to build a performant production-level cloud storage system~\cite{NSDI21Pangu}.

\noindent
\textbf{Generality of \sysname's techniques.}
\sysname's techniques can be applied to other  kinds of indexes.
Specifically, any lock-based indexes (e.g., bucket hash table) can use HOCL and command combination to improve concurrency performance and reduce round trips.
If an index follows lock-free search scheme, the two-level version mechanism is a good choice to mitigate write amplification.

% \input{secs/impl}
% \input{secs/discussion}
%\vspace{-0.1in}

\section{Evaluation}
\label{sec:eval}

In this section, we evaluate \sysname to answer following questions:
\begin{itemize}[
    leftmargin=*]

  \item How does \sysname perform under different workloads,
        and how do the different techniques employed in \sysname
        contribute to overall performance (\S\ref{eval:overall} and \S\ref{eval:range})?

        % \vspace{0.1cm}
  \item  How scalable is \sysname when varying client threads (\S\ref{eval:scal})?

        % \vspace{0.1cm}
  \item  How \sysname's techniques impact internal metrics, e.g., the number of round trips and write amplification (\S\ref{eval:micro})?

        % \vspace{0.1cm}
  \item How do parameters of \sysname, e.g., key size and cache index size, impact performance (\S\ref{eval:sen})?
  
  % \vspace{0.1cm}
  \item How does HOCL perform (\S\ref{eval:hocl})?

\end{itemize}

% Specifically, we first show the cumulative impact on 
% throughput and latency of various techniques under diverse workloads,
% and then present the scalability.
% Key components of \sysname are also evaluated.

\subsection{Setup}

\subsubsection{Hardware Platform.}
Since memory disaggregation hardware is unavailable,
we use a cluster of commodity, off-the-shelf servers to emulate MSs and CSs by limiting their usages of CPUs and memory~\cite{OSDI19LegoOS}.
Our cluster consists of 8 servers, each of which is equipped with
128GB DRAM, two 2.2GHz Intel Xeon E5-2650 v4 CPUs (24 cores in total), and one 100Gbps Mellanox ConnectX-5 NIC, installed with CentOS 7.7.1908 (Linux kernel version is 3.10.0).
All these servers are connected with a Mellanox MSB7790-ES2F switch.
For Mellanox ConnectX-5 NICs, the versions of driver and firmware
are ofed 4.7-3.2.9.0 and 16.26.4012, respectively.
Due to the limited size of our cluster,
we emulate each server as one MS and one CS.
Each MS owns 64GB DRAM and 2 CPU cores,
and each CS owns 1GB DRAM and 22 CPU cores.

\subsubsection{Compared Systems.}
FG~\cite{FG} is the only distributed \btree that supports disaggregated memory.
Since FG is not open-source, we implement it from scratch.
For fair comparison, we add necessary optimizations to it:
\begin{enumerate*}[label=(\itshape\roman*\upshape)]
  \item index cache for reducing remote accesses,
  \item using \rwrite to release lock rather than
  expensive atomic verb \rfaa.
\end{enumerate*}
In order to distinguish our modified version of FG from the original one,
we call it FG+ in the evaluation.
The performance of FG+ is higher than that reported in FG paper~\cite{FG}.

\subsubsection{Workloads.}
We explore different aspects of the systems by
using YCSB workloads~\cite{YCSB}.
We use five types of read-write ratio, as shown in Table~\ref{tbl:workloads}.
Note that insert operations include updating existing keys (about 2/3 of all insert operations).
\begin{table}[!t]
  \begin{center}

    % \resizebox{0.7\textwidth}{!}{
    \begin{tabular}{c|r|r|r}
      \hline
      Workload               & \ \ \ Insert \ \  & \ \ Lookup\ \  & Range Query \\
      \hline
      \hline
      \emph{write-only}      & 100\%             &                              \\
      % \hline
      \emph{write-intensive} & 50\%              & 50\%                         \\
      % \hline
      \emph{read-intensive}  & 5\%               & 95\%                         \\
      % \hline
      \emph{range-only}      &                   &                & 100\%       \\
      \emph{range-write}     & 50\%              &                & 50\%        \\
      \hline
    \end{tabular}
    % }

  \end{center}

  \caption{Workloads.}
  \vspace{-1cm}
  \label{tbl:workloads}

\end{table}

\begin{figure*}[!t]
  \centering

  \centering
  \includegraphics[width=\linewidth]{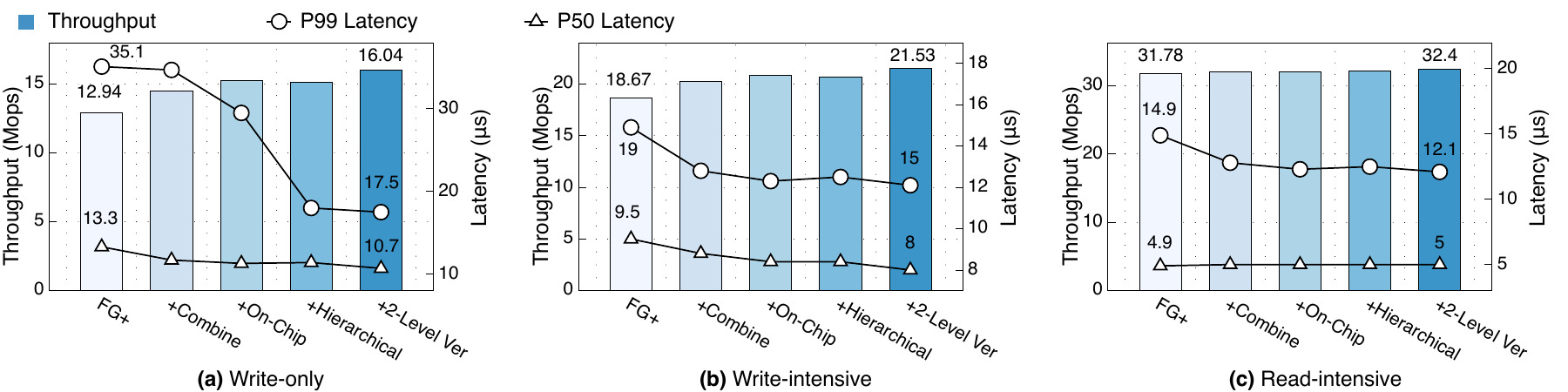}
  
  \vspace{-0.2cm}
  \caption{Contributions of techniques to performance
    (uniform workloads).}

  \label{fig:overall:uniform}
\end{figure*}

\begin{figure}[t!]

  \centering
  \includegraphics[width=0.9\linewidth]{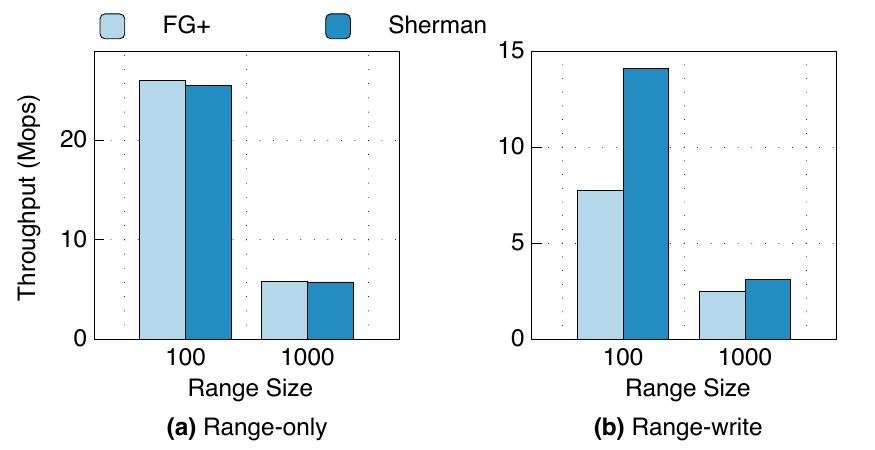}

  % \vspace{-0.1in}
  \caption{Performance of range query.
  }
  \vspace{-0.5cm}

  \label{fig:range}
\end{figure}

There are two types of key popularity: \emph{uniform} and \emph{skewed}.
In uniform workloads, all keys have the same probability of being accessed.
Skewed workloads follow a Zipfian access distribution
(Zipfian parameter, \ie skewness, is 0.99 by default),
which is common in production environments~\cite{YCSB, FAST20HotRing}.

Unless otherwise stated,
all the experiments are conducted with 8 MSs and 8 CSs.
Each CS owns 500MB index cache,
and launches 22 client threads (176 in total in our cluster).
For each experiment, we bulkload
the tree with 1 billion entries (8-byte key, 8-byte value) 80\% full,
then perform specified workloads.
The size of a tree node
(\ie internal node and leaf node) is 1KB.
All plots in this section present the average of 3 or more runs.

\subsection{Overall Performance}
\label{eval:overall}

To analyze \sysname's performance,
we break down the performance gap between FG+ and \sysname
through applying each technique one by one.
Figure~\ref{fig:overall:skew} and Figure~\ref{fig:overall:uniform}
show the results under skewed and uniform workloads, respectively.

In these two figures, \emph{\textbf{+Combine}} stand for command combination technique.
\emph{\textbf{+On-Chip}} and \emph{\textbf{+Hierarchical}} are two
design parts of HOCL:
leveraging on-chip memory of NICs to store locks,
and hierarchical structure with handover mechanism, respectively.
\emph{\textbf{+2-Level Ver}} represents two-level version mechanism,
and shows the final performance of \sysname.

\subsubsection{Skewed Workloads.}
\label{subsub:eval:skew}

We make following observations from Figure~\ref{fig:overall:skew}.
First, in both write-only and write-intensive workloads,
\sysname has much higher throughput and lower latency against FG+.
Specifically,
in write-only workloads,
\sysname achieves 24.7$\times$ higher throughput
with 1.2$\times$ lower median latency (50p latency)
and 35.8$\times$ lower 99th percentile latency (99p latency).
In write-intensive workloads,
\sysname achieves 23.6$\times$ higher throughput
with 1.4$\times$/ 30.2$\times$ lower 50p/99p latency.

Second,
all techniques contribute to
the high write efficiency of \sysname.
Here we analyze each technique
in terms of write-intensive workloads
(write-only workloads has the same conclusions):
\ding[1.2]{192} Command combination improves
the throughput by 3.37$\times$ and reduces
the 50p/99p latency by 1.14$\times$/3.18$\times$,
since it saves at least one round trip for each insert
operation (two round trips in case of some tree split events)
and further shortens the critical paths,
decreasing the probability that conflicting requests are blocked.
\ding[1.2]{193} By putting locks into on-chip memory of NICs,
\sysname gains 3.06$\times$ and 3.48$\times$ improvement
in terms of throughput and 99p latency, respectively.
This is because
on-chip memory avoids PCIe transactions of lock operations at MS-side,
which provides extremely high throughput for RDMA atomic verbs and
thus can absorb more failed \rcas for retries.
Moreover, since the PCIe transactions are excluded from
the critical path of \rcas,
conflicting \rcas commands experience shorter queueing time in NICs,
ensuring lower tail latency (\ie 99p latency).
\ding[1.2] {194}
Introducing hierarchical lock structure to
\sysname brings 2.22$\times$ higher throughput,
1.12$\times$ lower 50p latency and
2.68$\times$ lower 99p latency,
which comes from the following three reasons.
First,
by acquiring local locks before remote ones,
unnecessary \rcas retries from the same CSs
are avoided,
mitigating the consumption of RDMA NICs' limited IOPS.
Second,
the local locks in each CS guarantee first-come-first-served fairness
by maintaining wait queues, thus lowering tail latency
due to starvation.
Third, the handover mechanism saves one round trip opportunistically.
\ding[1.2] {195}
Two-level version mechanism does not bring considerable throughput improvement (only \%3), since the major bottleneck is
concurrent conflicts rather than RDMA IOPS at this time;
the 50p latency is reduced by 400ns (from 7.3$\mu$ to 6.9$\mu$),
since smaller \rwrite IO size has shorter PCIe DMA time at both
CS-side and MS-side.

Third, in read-intensive workloads (Figure~\ref{fig:overall:skew}(c)),
\sysname does not present considerable performance improvement,
as expected, since all techniques we proposed aims to boost
performance of write operations.
Yet, there are still two points worth noting here:
\ding[1.2] {192} By saving round trips for 5\% insert operations,
command combination reduces 99p latency from 15.3$\mu$s to 12.9$\mu$s.
\ding[1.3] {193} \sysname increases the 50p latency by 100ns (2\%),
we contribute it to unsorted leaf node layout,
which
causes traversal of the entire leaf node even for non-existing keys.

\begin{figure*}[!t]

  \centering
  \includegraphics[width=\linewidth]{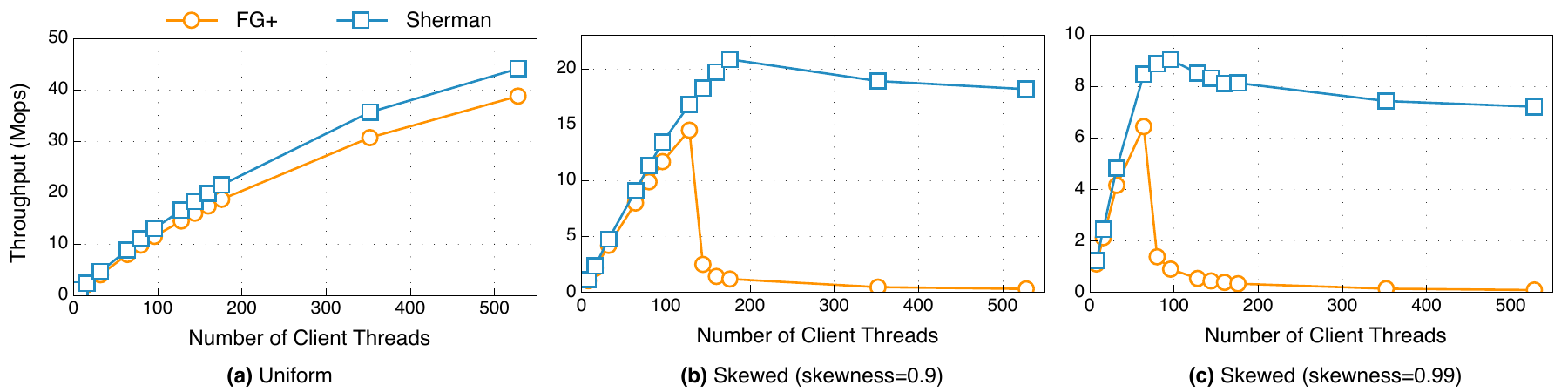}
  
  \vspace{-0.1cm}
  \caption{Scalability of \sysname (write-intensive workloads).}

  \label{fig:scale}
\end{figure*}

\begin{figure*}[!t]

  \centering
  \includegraphics[width=\linewidth]{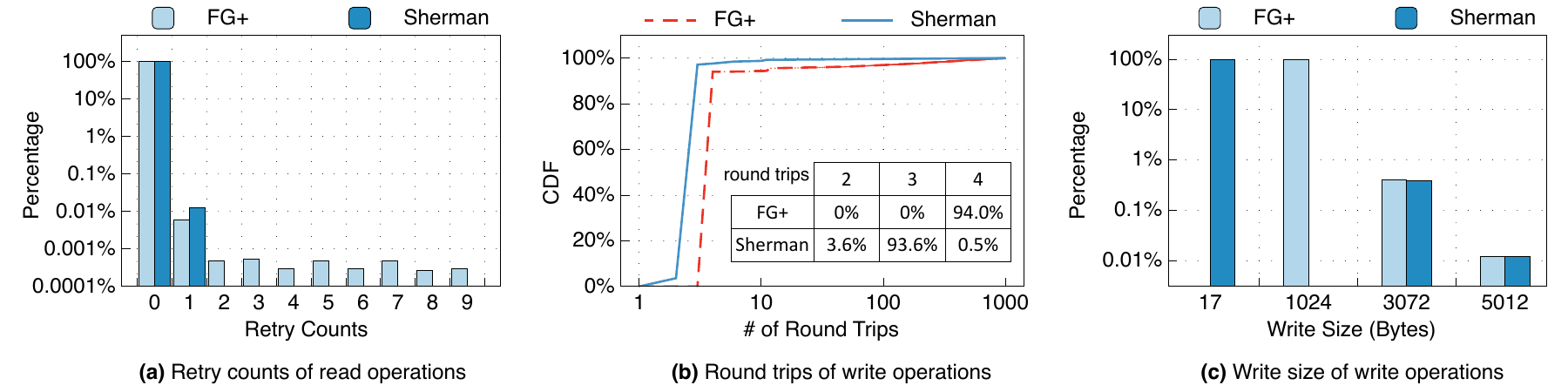}

  \caption{In-depth analysis using internal
    metrics (write-intensive workloads, skewness=0.99).}

  \label{fig:micro}
\end{figure*}

\subsubsection{Uniform Workloads.}

As shown in Figure~\ref{fig:overall:uniform}, compared with FG+,
\sysname delivers 1.24$\times$ and 1.15$\times$ higher throughput
in write-only and write-intensive workloads, respectively.
These improvements mainly come from command combination and two-level version.
Command combination saves round trips, so each client thread can execute
more insert operations per second;
two-level version reduces IO size of \rwrite from node size to
entry size, thus giving full play to
the RDMA's characteristics of extremely high small IO rate.
HOCL is designed for high-contention scenarios (\ie skewed workloads),
so it does not increase throughput in uniform workloads.
As for latency, \sysname reduces 50p/99p latency by 1.24$\times$/2.01$\times$
and 1.19$\times$/1.27$\times$ in write-only and write-intensive workloads, respectively, which mainly is contributed to command combination
and HOCL:
command combination saves round trips,
and HOCL saves PCIe transaction
time in MS-side as well as improves the fairness of locks.

\subsection{Range Query Performance}
\label{eval:range}

In this experiment, we evaluate the performance of range query by using range-only and range-write workloads. The targeted range follows the skewed access pattern.
Figure~\ref{fig:range}(a) shows the performance of range-only workloads,
from which we make two observations.
First, when the range size equals 100,
FG+ outperforms \sysname by 2\%.
This is because the
unsorted leaf layout in \sysname leads
to unnecessary scans when targeted
leaf nodes are partially occupied.
Second, as range size grows (i.e., 1000), the throughput of \sysname and FG+ drops and is almost the same,
since network bandwidth becomes the bottleneck.

Figure~\ref{fig:range}(b) shows the throughput of range query,
under range-write workloads.
Half of the client threads issue insert operations, and the other half issue range query operations.
\sysname outperforms FG+ by up to 1.82$\times$.
This is because \sysname's write operations save a considerable quantity of network resources for range query operations.
Specifically, HOCL significantly decreases the number of RDMA messages via lock handover and the hierarchical structure;
two-level version mechanism reduces the \rwrite IO size from node level to entry level.

\subsection{Scalability}
\label{eval:scal}

In this experiment, we test the scalability of \sysname
by varying the number of client threads that concurrently manipulate the tree.
Due to the limited CPU cores in CSs, \ie 22 $\times$ 8 in total,
we bind multiple coroutines to every core of CSs;
coroutines are implemented via Boost C++ library.
Each coroutine stands for a client thread, which issues insert/lookup requests;
a coroutine yields after initiating RDMA commands,
allowing other coroutines to do useful work.
We use write-intensive workloads.
Figure~\ref{fig:scale} shows the results under
uniform and skewed scenarios.
We make the following observations.

First, in uniform workloads,
both \sysname and FG+ can scale well.
In case of 528 client threads,
\sysname achieves 44 Mops, 1.14$\times$ the throughput of FG+.
The improvement mainly comes from two-level version mechanism:
by writing back leaf nodes in a smaller granularity,
more RDMA bandwidth are saved for \rread commands and lock operations.

Second, in skewed workloads,
a higher contention degree (\ie larger skewness value)
leads to lower peak throughput.
Specifically,
\sysname achieves 21 Mops peak throughput in case of 0.9 skewness,
which is 1.44$\times$ higher than that of FG+;
and 9 Mops in case of 0.99 skewness,
which is 1.4$\times$ higher than that of FG+.
This is because the more serious the contention,
the more likely concurrent write operations to be blocked,
degrading the peak throughput.

Third, in skewed workloads,
\sysname can provide sustainable throughput
when more client threads are added;
yet, FG+ experiences performance collapse.
We attribute the \sysname's stable performance
to a combination of
all techniques we proposed, as
stated in \S\ref{subsub:eval:skew}.

\begin{figure*}[!t]

  \centering
  \includegraphics[width=\linewidth]{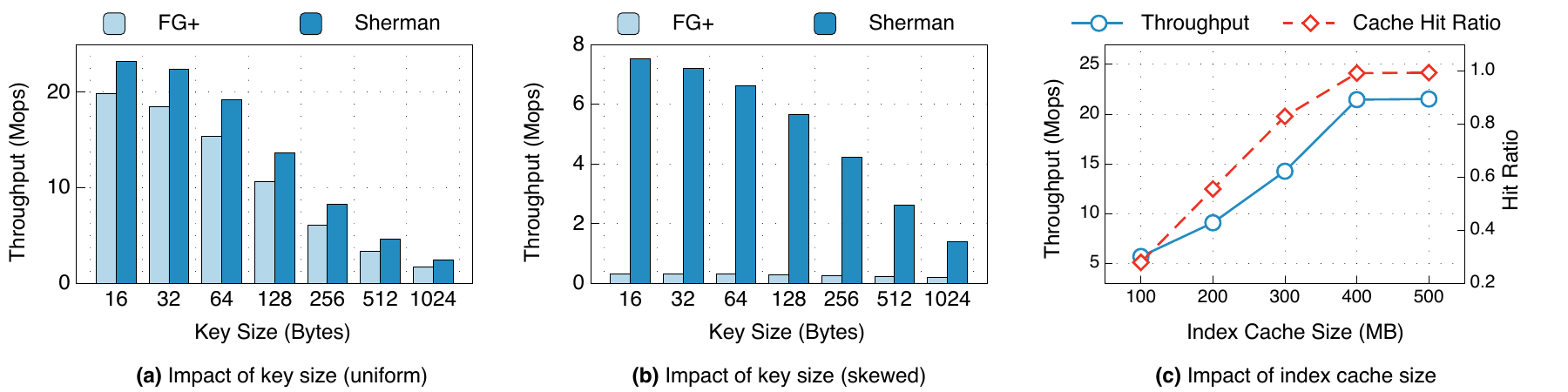}
  
  \vspace{-0.2cm}
  \caption{Sensitivity analysis.}

  \label{fig:sensti}
\end{figure*}

\begin{figure}[t!]

  \centering
  \includegraphics[width=0.9\linewidth]{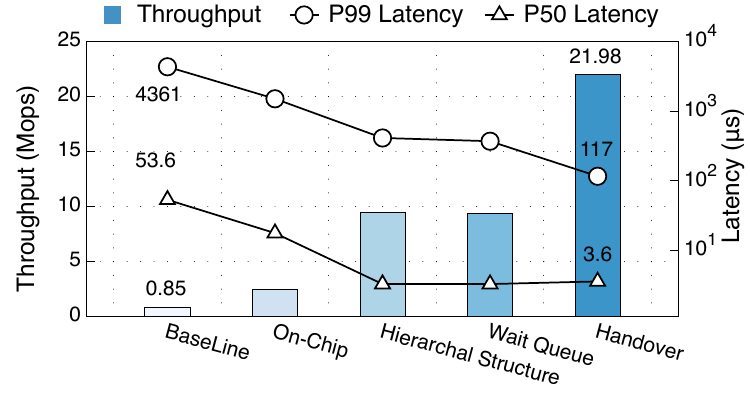}

  \vspace{-0.2cm}
  \caption{Performance of HOCL with skewed pattern.}

  \label{fig:hocl}
\end{figure}

\subsection{In-Depth Analysis}
\label{eval:micro}

In order to unveil detailed information of \sysname, we collect statistics for various internal
metrics, including the number of read retries, the number of round trips, and write size,
to perform an in-depth analysis.
We choose write-intensive workloads with a skewed access pattern (skewness=0.99).

\subsubsection{Retry Counts.}
Figure~\ref{fig:micro}(a) shows retry counts of lookup operations.
We make two observations.
First,
for FG+ and \sysname, 99.98\% of lookup operations do not requires retries.
This is because PCIe links already guarantee a certain level of atomicity:
a PCIe read transaction is strictly ordered after prior PCIe write transactions~\cite{SOCC20RDMAPM}.
Such a guarantee circumvents most interleaving accesses between read and write operations.
Second, FG+ experiences a few multiple times (e.g., 9) of read retries.
This is because FG+ does not adopt \sysname's two-level mechanism, causing larger RDMA IO size with longer DMA times at MS-side and thus increasing the likelihood that lookup operations fetch inconsistent tree nodes.

\subsubsection{The Number of Round Trips.}
Figure~\ref{fig:micro}(b) reports the cumulative distribution of write operations' round trips.
We make three observations.
First, 
94\% of write operations need 4 round trips in FG+, while 93.6\% need 3 round trips in \sysname.
The reason is that \sysname uses command combination technique to coalesce write-back and lock release.
Second, 3.6\% of write operations in \sysname only need 2 round trips, since HOCL's handover mechanism saves one round trip opportunistically.
Third, for FG+, the 99th percentile of round trips is 453, which explains why its tail latency is so high (\S\ref{subsub:eval:skew}).
In contrast, \sysname leverages HOCL to avoid massive lock retries and offer fairness; thus, its  99th percentile of round trips is 11, ensuring low tail latency.

\subsubsection{Write Size.}
Figure~\ref{fig:micro}(c) shows the write size of write operations.
Since we use skewed workloads which features strong access locality, 
only about 0.4\% of write operations trigger node splits, inducing large than 1KB writes.
As shown in the figure, for write operations without node splits, \sysname only needs to write back 17 bytes (i.e., 16-byte key-value pair along with two 4-bit versions) by using two-level version mechanism, rather than writing back the whole tree node, thus eliminating write amplification.

\subsection{Sensitivity Analysis}
\label{eval:sen}

\subsubsection{Key Size.}
Here we study how key size affects \sysname's performance.
Since \sysname embeds keys in tree nodes, 
in this experiment we fix the number of entries in a leaf node to 32 by changing the size of leaf nodes.
We bulkload the tree with 200 million entries 80\% full and then perform write-intensive workloads.
Figure~\ref{fig:sensti}(a) shows the throughput under uniform workloads.
We make two observations.
First, 
as the key size grows,
the performance of both \sysname and FG+ drops.
This is because every index operation needs to fetch the
whole leaf node via \rread,
and thus larger node size (caused by larger key size) consumes more network bandwidth for 
each index operation.
Second, 
as the key size grows from 16B to 1KB,
the performance advantage of \sysname over FG+ increases from 1.17$\times$ to 1.47$\times$, 
since two-level version in \sysname can save more bandwidth for larger tree nodes.

Figure~\ref{fig:sensti}(b) presents the throughput under skewed workloads.
Since FG+ suffers from low throughput in high-contention scenarios,
the increasing key size does not affect its throughput.
When the key size is 1KB, \sysname still outperforms FG+ by 1.4$\times$.

\subsubsection{Index Cache Size.}

In order to study how the index cache size affects the performance of \sysname,
we conduct an experiment with uniform and write-intensive workloads.
Figure~\ref{fig:sensti}(c) shows the result.
As the cache index capacity grows,
both \sysname's performance and cache hit ratio increase.
For large dataset (\ie 1 billion entries in our evaluation),
a 400MB index cache can bring a cache hit rate close to 98\%,
which demonstrates the efficiency of our index cache design.

\subsection{HOCL Performance}
\label{eval:hocl}

In this experiment, we evaluate the performance
of HOCL.
We launch 176 threads across 8 CSs
to acquire/release 10240 locks stored in an MS.
These operations follow a skewed access pattern with 0.99 skewness.
Figure~\ref{fig:hocl} shows the result.
Putting locks into on-chip memory
improves the throughput by 2.89$\times$ and
reduces the 50p/99p latency by 3.01$\times$/2.88$\times$,
since on-chip memory eliminates PCIe transactions for
\rcas commands at MS-sides,
making processing units in NICs more efficient and
shortening queueing time of conflicting commands.
By introducing local lock tables at CS-sides and further forming a
hierarchical structure,
we gain 3.85$\times$, 5.39$\times$, and 3.65$\times$
improvement in throughput, 50p latency and 99p latency,
respectively.
This is because
a thread can issue \rcas to acquire a remote lock
only when no thread at the same CS holds this lock,
avoiding a large amount of failed \rcas retries and
further saving RDMA IOPS.
Wait queues provide first-come-first-served fairness
within a CS;
as a result, the 99p latency is reduced from
414$\mu$s to 372$\mu$s.
Handover mechanism further improve
the throughput by 2.34$\times$ and
reduces the 99p latency by 3.19$\times$,
since by handing over locks from a thread
to another thread locally,
remote locking via \rcas commands can be avoided, accelerating lock acquisition.

%\vspace{-0.1in}
\section{Related Work}

To our knowledge, \sysname is the first tree index on disaggregated memory that 
can deliver high performance for both read and write operations with commodity RDMA NICs.
We discuss two aspects of related work: RDMA-based databases and memory disaggregation.

\subsection{RDMA-based Databases}
Fast RDMA network spurs researchers to build new distributed databases.
FaRM~\cite{FaRM, FaRMOpacity} provides general distributed transactions by using RDMA for messaging and direct-access to remote memory.
DrTM~\cite{DrTM} leverages the strong consistency between RDMA and HTM (hardware transaction memory)
to transform a distributed transaction into a local one.
FaSST~\cite{OSDI16FaSST} argues that two-sided unreliable verbs have higher scalability than one-sided verbs; thus, FaSST proposes a fast RPC framework using two-sided unreliable RDMA 
and builds an OCC-based distributed transaction engine on it.
Chiller~\cite{SIGMOD20Chiller} proposes a contention-centric partitioning scheme to
improve throughput of RDMA-based transactions.
Aurogon~\cite{FAST22Aurogon} redesigns timestamp ordering protocols with RDMA to reduce aborts.
Other databases exploit RDMA to support distributed join operations~\cite{SIGMOD15RDMAJOIN,VLDB17RDMAJOIN}.
These above systems can use \sysname to index data, or employ HOCL to boost their concurrency performance.

Other distributed database systems assume a NAM (network-attached memory) architecture that \emph{logically} decouples compute and memory servers and uses RDMA for communication~\cite{VLDB17NAMDB,VLDB16NAM, FG}.
Specifically, NAM-DB~\cite{VLDB17NAMDB} designs a scalable global counter technique to support 
snapshot isolation efficiently.
Compared with NAM architecture, memory disaggregation assumed by \sysname is more radical:
it decouples compute and memory \emph{physically}, contributing to an ideal independent scaling of compute and memory.
Furthermore, memory disaggregation poses more challenges to database design:
near-zero computation power at memory-side and small local memory at compute-side.

% Logging
Active-Memory~\cite{VLDB19ActiveMemory} leverages RDMA to replicate data in a failure
atomic way.
It uses ordering property of RDMA to combine undo logging and in-place update in a single network round trip.
Inspired by Active-Memory, \sysname combines RDMA commands within an index write operation to reduce round trips.

DFI~\cite{SIGMOD21DFI} abstracts low-level RDMA verbs and provides a set of flexible interfaces
(e.g., multicast, global sequencers) to support data-intensive applications such as OLAP.
It will be interesting if DFI exposes on-chip memory of RDMA NICs to applications,
to accelerate some interfaces such as global sequencers.

RDMA also accelerates indexing in distributed databases. As discussed in \S\ref{sec:moti},
most RDMA-based indexes do not support disaggregated memory.
FG~\cite{FG} is the first RDMA-based index that can be deployed on disaggregated memory,
since it uses one-sided verbs for all index operations.
RACE~\cite{ATC21RACE} is an extendible hashing index on disaggregated memory that supports lock-free accesses and remote resizing.
\sysname rethinks how to combine RDMA hardware features and RDMA-friendly software designs to
improve tree indexes' performance on disaggregated memory.

Finally, some researchers design general distributed shared memory (DSM) with RDMA~\cite{VLDB18GAM,FAST21Concordia, SOCC17Hotpot,TOS20THDPMS}, to support database components such as transaction engine.
Specifically, GAM~\cite{VLDB18GAM} provides a unified global memory abstraction and maintains coherence between cache copies via a directory-based protocol.
Concordia~\cite{FAST21Concordia} leverages programmable switches to mitigate cache coherence overhead.
Hotpot~\cite{SOCC17Hotpot} and TH-DPMS~\cite{TOS20THDPMS} support data durability by using persistent memory.
Index cache in \sysname exploits the structure of \btree to invalidate stale cached entries lazily without any coherence protocols.

\subsection{Memory Disaggregation}
Memory disaggregation is not a new idea:
Lim et al.~\cite{ISCA09Memory} proposed it more than a decade ago,
to attack the problem of a growing
imbalance in compute-to-memory-capacity ratio.
Recently, memory disaggregation regains attention for two reasons.
First,
large-scale companies report the low memory utilization in datacenters~\cite{OSDI19LegoOS, IWQoS19Trace, EuroSys20Borg}.
Second, high-speed RDMA network makes performance of remote accesses close to that of local accesses.
Gao et al.~\cite{OSDI16Disaggregation} find that, under a memory disaggregation architecture, 40-100Gbps network bandwidth and 3-5$\mu$s network latency can maintain application-level performance.

Many recent academic efforts have been devoted to making memory disaggregation practical.
LegoOS~\cite{OSDI19LegoOS} designs a distributed operating system to manage disaggregated resources.
pDPM~\cite{ATC20pDPM} explores how to deploy disaggregated persistent memory (PM) in an efficient manner.
Rao et al.~\cite{ANCS16SQL} demonstrate that, for Spark SQL, memory disaggregation is promising with currently available network hardware.
Zhang et al.~\cite{VLDB20Disaggreation} evaluate two production DBMSs, PostgreSQL and MonetDB, on disaggregated memory using TPC-H benchmark;
they find that in several scenarios, memory disaggregation can boost performance of databases.
LegoBase~\cite{VLDB21LegoBase} and PolarDB Serverless~\cite{SIGMOD21PolarDB} co-design
database and memory disaggregation, achieving faster failure recovery speed than traditional architectures.
\sysname focuses on building a fast \btree index on disaggregated memory.

Industry is proposing memory disaggregation hardware.
HPE's "The Machine"~\cite{TheMachine,keeton2015machine} 
connects a set of SoCs to a shared memory pool via photonic network.
IBM's ThymesisFlow~\cite{MICRO20ThymesisFlow} 
leverages OpenCAPI~\cite{OpenCAPI} to enable CPUs directly access remote disaggregated memory without software involvement.
VMware uses cache coherence mechanisms to track applications' memory accesses, reducing data amplification of disaggregated memory~\cite{ASPLOS21DisMem}.
%\vspace{-0.15in}
\section{Conclusion}
We proposed and evaluated
\sysname, an RDMA-based \btree index on disaggregated memory.
\sysname introduces a set of techniques
to boost write performance and 
outperforms existing solutions.
We believe \sysname demonstrates that
combining RDMA hardware features and 
RDMA-friendly software designs can enable 
a high-performance index on disaggregated memory.

% \pagebreak

% \pagebreak

\bibliographystyle{ACM-Reference-Format}
\bibliography{paper}

\end{document}